\documentclass[sigconf,10pt]{acmart}

\usepackage[utf8]{inputenc}

\graphicspath{{graphs/}}
\usepackage{float}
\usepackage{array}
\usepackage[]{subfig}

\setcopyright{none} 
\acmConference[IMC '19]{Internet Measurement Conference}{October 2019}{Amsterdam, Netherlands}
\acmYear{2019}

\begin{document}

\title{{\huge Measuring Irregular Geographic Exposure on the Internet}}

\author{Jordan Holland}
\affiliation{%
  \institution{University of Tennessee, Knoxville}
  \institution{Princeton University}}
\email{jordanah@princeton.edu}
\authornote{Jordan completed this work while at the University of Tennessee, Knoxville.}

\author{Jared M. Smith}
\affiliation{%
  \institution{University of Tennessee, Knoxville}}
\email{jms@vols.utk.edu}

\author{Max Schuchard}
\affiliation{%
  \institution{University of Tennessee, Knoxville}}
\email{mschucha@utk.edu}

\begin{abstract}
    We examine the extent of needless traffic exposure by the routing infrastructure to nations
    \emph{geographically} irrelevant to packet transmission. We
    quantify what countries are \emph{geographically logical} to observe
    on a network path traveling between two nations through the use of
    convex hulls circumscribing major population centers. We then
    compare that to the nation states observed in over 2.5 billion
    measured paths. We examine both the entire geographic topology of
    the Internet and a subset of the topology that a Tor user would
    typically interact with. We reveal that 44\% of paths concerning the
    entire geographic topology of the Internet and 33\% of paths in the user experience
    subset unnecessarily expose traffic to at least one nation, but often more. 
    Finally, we consider the scenario where countries exercise both legal and physical 
    control over autonomous systems, gaining access to traffic outside of their
    geographic borders, but carried by organizations that fall under the AS's registered
    country's legal jurisdiction. At least 49\% of paths in both measurements expose
    traffic to a geographically irrelevant country when considering both
    the physical and legal countries that a path traverses.
\end{abstract}



\maketitle

\section{Introduction}
\label{sec:intro}

The Internet is comprised of independent networks called Autonomous
Systems (ASes) which depend on each other for inter-network
connectivity. Network traffic often traverses multiple ASes in
order to reach its final destination. Any adversarial transit AS
situated between sender and receiver, termed a \emph{path based
  adversary}, can degrade network availability, violate data
integrity, and undermine confidentiality. While a single malicious
transit AS is a powerful adversary, nation states represent an even
stronger path based adversary. A nation state has the ability to
exercise control over both ASes that physically operate networking
infrastructure within its borders and ASes whose corporate governance
falls within the nation state's legal jurisdiction. A motivated
nation state could coerce multiple ASes into acting as colluding path
based adversaries on the nation state's behalf.

Revelations in recent years have demonstrated the extent to which
nation states are willing to exert pressure on private entities in an
effort to execute national cyber policy on network traffic. For
example, the United States, Great Britain, and other members of the so
called Five Eyes intelligence alliance have integrated dragnet
surveillance into core Internet transit links that reside within their
borders~\cite{derQuantum, wiredATT, arsATT}. Additionally, there
exist recorded instances of censoring nation states applying domestic censorship policies to network traffic that neither
originates nor is bound for domestic sources~\cite{chinaDNS, dnsOdd}. 

Due to the potential security concerns exposure to additional nation
states might present, it is reasonable to expect that the Internet's
routing infrastructure minimizes such exposures. In scenarios where
sender and receiver reside inside the same or adjacent nation states,
one would expect that these are the only jurisdictions traffic is
exposed to. Even in scenarios where the sender and receiver are
located in non-adjacent nations, and exposure to third party nation
states is a physical necessity, one might assume that exposure is
limited to a minimal set of nations required to build a physical link
between sender and receiver. However, routing decisions focus on
the \emph{logical} network topology rather than the \emph{geographic}
or \emph{political} topology. This can result in paths exposing
traffic to nations which do not lie between the geographic locations
of the sender and receiver. This excess exposure needlessly increases
the power of certain nation states if they elected to coerce ASes into
serving as path-based (i.e. on-path) adversaries on their behalf.

In this work, we examine to what extent the Internet's routing
infrastructure increases the capacity of certain nation states to
undermine security properties of network traffic in excess of what would be
predicted. Specifically, we quantify how often network traffic is
exposed to additional nation states beyond those residing along a
\emph{geographically logical} path between sender and receiver. Building the set of geographically logical countries between sender and receiver is done with a novel technique based on computing a population biased convex hull between the sending country and receiving country. We define the set of \emph{geographically normal} countries providing transit between source and destination to be the set of all countries that lie at least partially in this convex hull.
We examine two populations of paths in this work. 
First, we examine a sample of the entire topology of the Internet, which we term \emph{all-to-all}. 
Second, we examine a subset of paths to the Alexa Top 1000~\cite{Alexa} through
the Tor browser, which we call our \emph{user experience} dataset. 
To explore the \emph{physical exposure} of data to nation states we compare the countries where network infrastructure observed on the paths is physically located to the set of geographically normal countries we would expect. 
We  also explore what happens when nation states, in addition to taking
advantage of physical exposure, exercise control over
ASes whose corporate governance operates within their legal
jurisdiction, and leverage that pressure to observe all data handled
by such ASes, which we term \emph{legal exposure}.

All together, we examine over \textbf{2.5 billion paths} from traceroutes conducted by CAIDA's Ark measurement framework~\cite{ark} and the RIPE Atlas measurement framework ~\cite{ripeAtlas}.
Overall, we find that 44\% of examined all-to-all paths and 37\% of the examined user experience paths physically expose traffic to \emph{at least} one unexpected nation state. Furthermore, we find that the quality of paths varies greatly on a country-to-country basis. As an example, more than half of the countries observed saw unexpected nation states on 80\% or more of their inbound paths.
Lastly, when we consider the countries that paths both legally and physically traverse, at least 49\% of paths in both datasets expose traffic to at least one unexpected nation state.

The remainder of this paper is presented as follows. In
Section~\ref{sec:background} we will cover relevant background on
how logical routing decisions are made, and expand on the
motivation for our study. In Section~\ref{sec:method} we will present
our methodology for collecting paths, labeling countries that the path
either physically or legally exposes data to, and lastly how we build
a quantifiable measure of what countries are geographically logical to
observe between source and destination. In Section~\ref{sec:dataset}
we will cover the properties of our resultant datasets.
Section~\ref{sec:physicalExposure} examines the extent to which
nations physically expose their traffic to other countries, along
with how often that exposure is geographically illogical.
Section~\ref{sec:legalExposure} expands this analysis to additionally
consider exposure to entities that could be legally coerced by a
particular country. Section~\ref{sec:related} compares our work to similar studies. Lastly we draw our final conclusions in
Section~\ref{sec:conc}.

\section{Background and Motivation} \label{sec:background}

\subsection{Path Selection On the Internet} \label{ssec:backRouting}

The Internet is comprised of a collection of independently
administered networks called Autonomous Systems, or ASes. ASes
provide network connectivity to hosts inside of their network,
enabling those hosts to connect to other host located either inside
the same AS or in a remote AS. To this end, ASes deploy special
purpose networking equipment called routers, whose job it is to
compute the best path between data sender and receiver, and forward
data to the next hop (router) in that path. In order to compute best
paths, routers execute routing protocols, which facilitate the
exchange of topology information between routers using that
information to compute best paths. Routing protocols do not compute
the best path to every individual IP address, rather, routing
protocols compute best paths to blocks of IP address, and forward
traffic addressed to any host inside that block along the same path.
Routing protocols are typically divided into two categories, inter-AS
and intra-AS routing protocols. Inter-AS routing protocols compute
the sequence of ASes data will travel when moving between remote ASes.
Intra-AS routing protocols compute the best path traffic takes inside
a given AS. Intra-AS routing protocols are responsible for both
getting traffic to a end destination inside the current AS and
delivering traffic to a boarder router that will transfer traffic to
the next AS along a multi-AS path.

The Border Gateway Protocol~\cite{RFC4271}, or BGP, is the de-facto
standard inter-AS routing protocol. The single standard routing
protocol is a result of the demand for interoperability between
independently managed organizations that often compete with each
other. BGP is a path vector routing protocol with policies. The
policy portion of BGP allows network administrators to select paths
based on arbitrary criteria, rather than simple the shortest paths.
Commonly, ASes utilize their business relationship with neighboring
ASes to make a first pass routing decision. ASes that have direct
connectivity with each other typically form customer-provider
relationships, where the customer pays the provider for all traffic
flowing between the two ASes in exchange for connectivity to remote
ASes via the provider. ASes will generally follow a routing policy
termed ``Valley Free Routing'', where they prefer routes that are more
economically advantageous, however recent measurement studies have
shown that this is not always the case~\cite{anwar2015investigating}.

Because intra-AS decisions only need to be computed over
infrastructure held by one organization, removing the demand for
interoperability, there are a myriad of intra-AS routing protocols
deployed. Examples include link state protocols such as
IS-IS~\cite{RFC1142} and OSPF~\cite{RFC2328}, along with path vector
variants such as EIGRP~\cite{RFC7868}. Most intra-AS protocols
include the ability to include network policy rather than simple
network distance as part of the routing decision making process. This
policy is often expressed as an ``administrative distance'' giving the
algorithm hints as to the administrator's preferences.

\subsection{Measuring Utilized Paths}  \label{ssec:traceroute}

Predicting the exact path data will travel between source and
destination is challenging. In the case of inter-domain routing, AS
relationships are closely held secrets. These relationships can be
inferred with some degree of accuracy based on publicly available
BGP mirrors, sources of information on the current state of the BGP
global routing table. However, as shown most recently by Anwar
et. al.~\cite{anwar2015investigating}, using inferred relationships to
predict inter-AS routing paths can be inaccurate. Predicting
intra-AS routes is even more challenging. First, \emph{which}
intra-AS routing protocol an AS is using is a corporate secret, and
difficult to detect. Second, the particular configuration and
administrative preferences that factor into the intra-AS routing
protocol's decision making process are difficult to both infer and to
use as a predictive model accurately.

An alternative, but more accurate, method for determining the utilized
path between hosts is to execute a {\tt traceroute}~\cite{RFC1393}
between the two hosts. Traceroute takes advantage of how networking
equipment commonly behaves when it encounters a packet with an expired
Time-To-Live (TTL) field. Often, but not always, routers will respond
with an ICMP message to the sending host, informing the host that the
packet expired, and the IP address at which the packet expired. By
sending packets with small, but incrementally increasing, TTLs, and
recording the IP addresses that respond, a host can map the sequence
of routers a packet traverses on its path. While traceroute provides a high
level of accuracy, there are several shortcomings. Most obviously,
the source must be under the control of the measuring party. To
overcome this shortcoming, there are several distributed measurement
test-beds that either conduct traceroutes at regular intervals to large
portions of the Internet or conduct traceroutes to specified
hosts. Examples of such test-beds include CAIDA's Ark
Infrastructure~\cite{ark} and RIPE's Atlas
Infrastructure~\cite{ripeAtlas}. Another major issue with traceroute is that network
infrastructure is not \textit{obligated} to respond when a packet's TTL expires. 
In this case gaps in the full path to the host will result.

\subsection{Path Based Adversaries} \label{ssec:pathAdv}



The security properties of many distributed systems can be impacted by
the adversarial AS that transits data, what we refer to as a \emph{path
  based adversary}. As an example, a path based adversary can
trivially violate the confidentiality of any unencrypted traffic.
Despite this obvious threat vector, a recent study by the EFF found
that only about half of web traffic is actually
encrypted~\cite{effEncrypt}. More complex attacks undermining
confidentially are also possible. Consider an AS
that wishes to attempt to de-anonymize users of Tor, an anonymous
communication system. Feamster and
Dingledine~\cite{feamster2004location} showed that a path based adversary AS could undermine the
anonymity properties of the Tor network when
when the AS appears on a path between a user and their entry into the Tor
network, as well as appearing between their traffic's exit point from
the Tor network and its final destination. The integrity of distributed
systems can also be disrupted by adversaries who lie along a utilized
path. Apostolaki et. al.~\cite{apostolaki2016hijacking} demonstrated
that adversaries capable of observing traffic between 900 IP blocks
could control and edit interactions between a majority of the
computational power dedicated to mining Bitcoin, opening up the
possibility of double spending via forced forking of the blockchain.
Additionally, in 2014 attackers utilized compromised BGP speakers in an
effort to hijack communications between Bitcoin miners and their pool
servers~\cite{bitcoinTheft}, resulting in the theft of Bitcoins.

There exist both academic studies and real world examples of adversaries that can
control multiple ASes, becoming exceptionally wide reaching path based
adversaries. Johnson et. al.~\cite{johnson2013users} first
expanded the AS level Tor adversarial model to include such powerful
adversaries when they explored the capacity of Internet Exchange
Points to de-anonymize Tor users. Revelations from whistle blowers
including both the Snowden leaks~\cite{derQuantum} and earlier
revelations by the AT\&T contractor Mark Klein~\cite{arsATT, wiredATT}
demonstrated the willingness of the NSA and other spy agencies such as
GCHQ to integrate surveillance devices, and even systems which
actively inject data into network streams, inside core network
infrastructure located within their respective nations. In addition
to the attacks outlined above, documents have revealed a complex
infrastructure for violating user confidentiality by building
relationship graphs based solely on linking data senders and receivers.
Furthermore, there exists evidence of censoring nation states, for
example China, either accidentally or intentionally applying
censorship policies to traffic that neither originates in nor is bound
to domestic hosts~\cite{chinaDNS, dnsOdd}.


\subsection{Motivation}{\label{motivation}}

From a security perspective, exposing traffic to any unnecessary nation is
potentially problematic since it grants the opportunity for those nations to act
as path based adversaries. Examples of potential malicious activity includes
more than the aforementioned areas of recent work. More broadly, it includes
dropping the data, eavesdropping on the data, and even changing the contents of
the data, thus removing any expectation of data integrity. While exposure to
some nations is unavoidable, any traffic exposed to extraneous nations, i.e.
nations not physically necessary for the propagation of traffic, needlessly
increases the aforementioned security risks. \textit{Our goal in this paper} is
to algorithmically measure the fraction of paths that expose network traffic to nations that are geographically illogical, and to quantify how much extraneous traffic a nation state gets to observe and control from these geographically illogical paths. Geographically illogical paths occur both under \textit{physical} and \textit{legal} conditions.

There are two scenarios in which data can fall under the legal jurisdiction of a nation. The first, scenario is when  the data is handled by networking equipment physically located in the borders of a nation during transit. The second, less obvious scenario, is when data "legally" transits a nation. In this scenario a nation may have legal jurisdiction of data that does not physically transit a country due to an AS being legally registered in one nation while having physical infrastructure in another. This scenario is of particular interest as of late with the passing of the CLOUD Act in the United States stating that the United States may, in some scenarios, have access to data that is physically stored in other countries but under legal jurisdiction of the United States \cite{CloudAct}. This \textit{legal vs. physical} router presence motivates our work beyond purely adversarial tampering, particularly from a policy perspective.   
The ideal geographic situation for data traveling from nation A to nation B is a path that consistently moves \emph{towards} nation B. In other words, we would logically expect the geographic network level paths between two nations to approximately, but not exactly, traverse the shortest path between those nations. While this definition is simplistic, it does highlight certain path selection choices that are illogical from a geographic standpoint. For example, a path that goes the opposite direction that its destination should be considered illogical. Additionally, data with a destination that lies within the same nation it originated from should \emph{almost never} leave that nation.

\section{Experimental Methodology} \label{sec:method}

In order to quantify the amount of geographically illogical paths and which countries can exert additional control over network traffic we need to build two datasets. 
First, we must establish a set of countries that we would expect traffic to be logically exposed to during transit
between a particular source/destination pair based on geographic
realities. We term countries inside of this set \emph{geographically
  normal} for a given source and destination. Second, we must
establish the actual paths data takes from sources to destinations and
establish which countries the network traffic is exposed to. We compute both
which countries physically host the network infrastructure traveled,
termed \emph{physical exposure}, and which countries can exert legal
pressure on the ASes appearing along the path, termed \emph{legal
  exposure}. By comparing the set of countries a path exposes network traffic
to with the set of geographically normal countries, we can label the
path as either \emph{normal}, no excess exposure, or \emph{non-normal}, containing geographically illogical countries. We can also
quantify the number of times a country is a \emph{benefactor} of a non-normal
path: specifically the number of additional paths 
it can observe as the result of geographically irregular paths.

\subsection{Geographically Logical Paths} \label{ssec:geoLogicConstruction}

Defining the geographically normal countries for a source and destination
pair was done using a novel technique based on \emph{convex hulls}.
The \emph{convex hull} of a set of points $S$ in $n$ dimensions is
the intersection of all convex sets containing $s$. For $n$
points $p1, ..., pN$ the convex hull $C$ is then given by
the expression:
\begin{equation}
C = \sum_{j=1}^{N} \lambda_j \rho_j : \lambda_j \geq 0\ \forall j\ and\ \sum_{j=1}^{N} \lambda_j = 1  
\end{equation} \label{eq:1}
A more intuitive way to think about the definition of a convex hull: given a
set of points, imagine the shape a stretched rubber band
takes when encompassing all of them. 

Using Equation 1, we can build convex hulls
containing \emph{both} the set of points that define the country
containing the source and a set of points that define the country
containing the destination. These convex hulls are computed taking
the spherical nature of the Earth into account. Note that when a country is the 
source and destination of network traffic, we \emph{only} 
consider that country as normal and do not build a convex hull for the single country.
This convex hull construction efficiently
defines all points that lie between source and destination countries.
Source and destination were considered at the granularity of nation
states to mitigate limitations in the accuracy of GeoIP location
usage later. It should be noted that this coarser granularity only
\emph{increases} the number of countries considered geographically
normal, thus providing an estimate of geographically normal that tends towards logical over illogical.
We then compute the set of geographically normal countries by
finding all countries that either fully or partially fall within this convex hull. In order to detect countries that reside
inside the convex hull, we test if any of the 15 largest cities in the
given country resides inside the convex hull. To detect countries
that lie only partially inside of the convex hull, we test if any
point along the edge of the convex hull lies inside the borders of a
particular country. We consider the entire country to be geographically normal
if any of the 15 largest cities inside of a country or the border of a country
lies within the defined convex hull. 

One option for defining the set of points that make up a country is to
utilize the nation's political borders, provided in the Matic shapefile
dataset~\cite{shapefiles}. In order to test this approach, we
utilized shapefiles which contain points that define polygons of the
actual borders of each country. This approach tends to result in
convex hulls between two nations that contain countries which do not
lie in the path between those nations. One factor contributing to
this is countries with non-contiguous territories or remote
territorial holdings. An example of this is the United State's
convex hull when including Alaska, Hawaii, Puerto Rico, Guam, and other remote
territories, as the resulting convex hull covered roughly one quarter
of the earth's \emph{total} surface area. Additionally, the political
borders of a country do not necessarily reflect where bulk the
Internet infrastructure of the country is located; as this generally
lies in the more populated areas. Relevant examples of this include
China and Russia. To address, this we built a \emph{separate} definition of
points describing a country using the latitude and longitude of the top 15 most populous cities in that country~\cite{pop}.

\begin{figure}
    \centering
    \includegraphics[width=0.75\columnwidth]{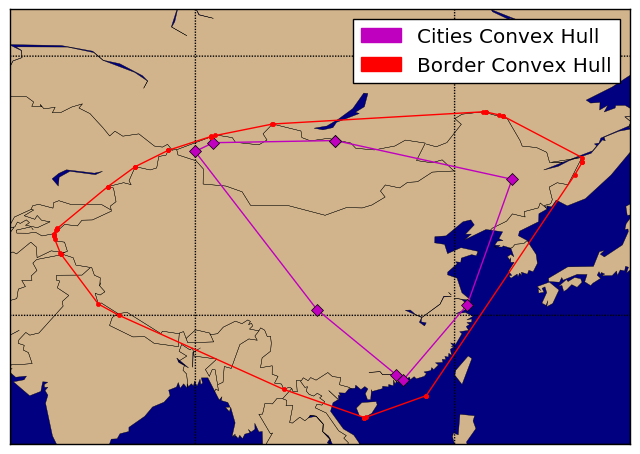}
    \caption{A comparison of the border based convex hull and
      population biased convex hull between China and Mongolia. Note
      that over 83\% of China's population lives on its
      eastern coast.} \vspace{-10pt}
    \label{fig:convex-hull}
    \vspace{-10pt}
\end{figure}

Figure~\ref{fig:convex-hull} shows an example of the two construction
techniques for a selected path between China and Mongolia. The population-based
convex hull results in a stricter version of geographically normal
between two countries and accurately reflects the fact that 83\% of
China's population, including all of its major cities, reside in the
eastern portion of the country. The border based convex hull includes
countries in the wrong cardinal direction, such as India and Vietnam,
a result of China's concave shape. We \emph{chose to use the city or population based
construction of a convex hull} for the measurements contained inside
this work. When a country is \textit{both} the source and destination of traffic
we consider traffic \textit{not normal} if any other country is found on
the path from source to destination. 

\subsection{Building Path Datasets} 

Our goal is to measure the physical and legal exposure of paths on the Internet.
Ultimately, we do this from two perspectives: an all-to-all perspective and a
user experience perspective. In measuring the exposure of paths from an
all-to-all perspective, we attempt to gain insight into the entire geographic
topology of the Internet. We realize, however, that many users only interact
with a subset of the entire topology due to content being cached at CDNs that
are geographically local to the end user. Therefore, it is important to separate
and compare these two perspectives to understand the similarities and
differences between the geographic topology that a user sees and what is
actually available. We take advantage of two commonly used traceroute frameworks
to obtain a statistically significant sample size for each perspective.

\subsubsection{All-To-All Path Set}

\begin{figure}[]
\centering
\begin{minipage}[t]{0.75\columnwidth}
  \centering
  \includegraphics[width=\columnwidth]{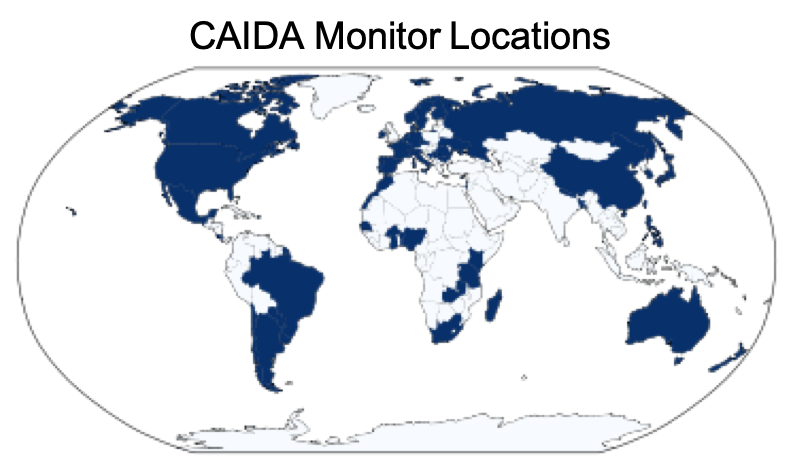}
  \caption{CAIDA's monitor locations make it ideal for measuring the all-to-all geographic topology of the Internet.}
    \label{fig:caida-monitor-locations}
\end{minipage}%
\hfill
\begin{minipage}[t]{0.75\columnwidth}
\centering
  \includegraphics[width=\columnwidth]{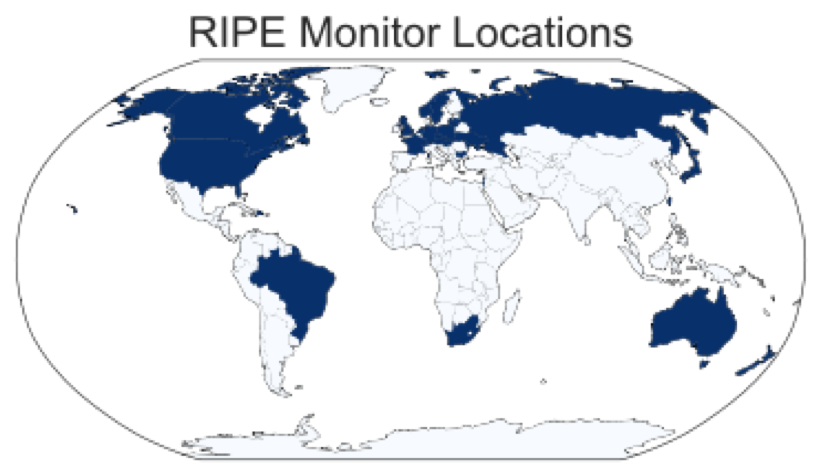}
  \caption{Our selected RIPE Atlas  probe/Tor proxy pairs cover every continent like CAIDA, though less countries are represented.}
    \label{fig:ripe-monitor-locations}
\end{minipage}
\vspace{-10pt}
\end{figure}

First, we leverage CAIDAs Archipelago Measurement Infrastructure (ARK)~\cite{ark}. ARK allows us to measure the entire geographic topology of the Internet. Comprised of over 180 monitors, ARK coordinators monitors which work in teams to send traceroutes to a random IP address in each block of globally addressable IPs~\footnote{Specifically, each announced /24 prefix.} Note that this does not imply that each monitor probes every announced prefix every 48 hours, but that a single destination prefix will be probed by \emph{one} monitor each cycle. There are two important features that make this measurement system the best for measuring the geographic topology of the Internet as a whole. First, it provides a uniform sample across the set of possible destinations by probing each announced /24 prefix. Second, it gives us both a geographically and legally diverse monitor set to probe from. The ARK measurement system has probes in over 160 ASNs and 64 unique countries. Figure \ref{fig:caida-monitor-locations} shows CAIDA monitors located on every major continent.

\subsubsection{User Experience Path Set}
The majority of (source, destination) pairs on the Internet are not selected in an "all-to-all" fashion. Users typically access content cached in Content Distribution Networks, or CDNs~\cite{labovitz2010internet}. CDNs replicate content, storing content on several servers scattered across the Internet. When users attempt to access content stored in a CDN, they are typically directed to the CDN server that is closest to them in terms of the network topology. This has two impacts on the paths user data typically travels over. First, in general we expect paths from users to the CDN nodes serving them to be shorter than the average path length on the Internet, due to the emphasis on content locality. Second, two users accessing the same piece of content might fetch it from different servers, depending on where the users are located. In order to build an accurate model of paths users take to popular content, we need to establish what servers users in a particular network connect to when accessing web content and what the path that specific user would take to those servers.

We visited the Alexa Top 1000~\cite{Alexa} from a collection of geographically
distributed vantage points to measure what IP addresses are accessed when users
load popular web content. We then recorded the remote hosts connected to in the process of fetching the content. Websites were loaded using a full version of the Chrome browser, driven by automation provided by the Selenium~\cite{Selenium} framework. It is vital that websites be fetched using a full browser, rather than simply using DNS resolution of the domain or \texttt{wget}, as a large fraction of the content on websites is loaded only when the CSS and Javascript on a page are processed by a browser rendering engine. The page loads are repeated from a geographically diverse collection of locations because the location a server's content is fetched from for a particular user vary based on the user's location, recall the note on CDN locality.

Page loads were conducted through the Tor Network~\cite{Tor}, an anonymous communication tool, which has open proxies scattered around the globe. We use Tor because one of our primary motivations is to understand how users could be affected by censorship from Geographically-Illogical paths. Since Tor is the de facto Censorship-resistant browser, we chose this as our HTTP proxy for Selenium. All DNS resolutions during this data gathering phase were conducted through available public Tor proxies to ensure that our crawler was pointed to the resources that were appropriate for the location of \emph{the Tor proxy}. To ensure that the Tor browser was not hitting CAPTCA pages and in fact loading all homepage content, we conducted an analysis of the page content through Tor. We found when using a standard web framework analysis tool, Wappalyzer~\cite{Wappalyzer}, that only 34 of the top 1000 sites contained reCAPTCHA scripts on their homepages. While these reCAPTCHA versions may not have blocked the content in our measurement, we point out that this content was only detected in 3.4\% of the top 1000 pages.

We take advantage of the RIPE Atlas~\cite{ripeAtlas} measurement ecosystem to
measure the path a user's communication would travel to and from these servers.
RIPE Atlas allows network operators to conduct measurements to user defined IP
addresses from a set of geographically distributed probes. During our crawling
of the Alexa Top 1000, we only selected Tor proxies that were located in the
same AS/country pair as at least one RIPE Atlas node. For each crawl of the
Alexa Top 1000, we conducted traceroutes from the RIPE Atlas probes to all IP
addresses observed during page loads conducted from that particular AS/country
pair. This set of traceroutes is our "User Experience" dataset.

\subsection{Computing Exposure}\label{ssec:computingexposure}

The datasets built in the prior section provide us two collections of traceroutes representative 
of particular populations of paths found on the Internet. We must map each IP
address found in these paths to the countries that the data is physically and
legally exposed to during transit in order to establish if these paths expose
traffic to extraneous countries.

\subsubsection{Computing Physical Exposure}

Computing physical exposure of a traceroute involves mapping each IP address in
the traceroute back to the country it is physically located in. We must use one
of many IP geolocation services that exist to solve this problem. There are multiple issues with these services. 
First, many are not free, or free at scale, so any IP queries to them must be carefully spent. 
Second, while the services are historically good at geolocating end hosts (at a country granularity), 
they have been known to be less precise when attempting to geolocate routers~\cite{netacuity}. We chose to use NetAcuity to geolocate our IP addresses, as research has recently shown them to be \emph{best} at geolocating routers~\cite{digitalelement}. We do not consider the goelocation of any IP address at a
resolution finer than the country level to limit the effects of IP geolocation
on our results.

\subsubsection{Computing Legal Exposure}

Computing the legal exposure of a tracreoute is a more exact science. We map
each IP address in the path back to the AS who owns the IP address by consulting
snapshots of BGP routing tables from RouteViews~\cite{routeviews}. More
specifically, we determine the AS who owns an IP address by examining who
originated the path for the block of the IP addresses \textit{on the day the
traceroute was conducted}. We then map the AS to its legal
jurisdiction with country data from the IANA registry~\cite{iana}.


\subsection{Common Processing Pipeline}

We built a common pipeline to process each traceroute from the all-to-all and
user experience datasets using the information gathered from Section~\ref{ssec:computingexposure}. Due to the monetary cost of geolocating IP
addresses, we have carefully chosen how we have spent our queries to NetAcuity.
Specifically, we geolocate every router found in the CAIDA traceroutes. Note
that this does not mean we have geolocated every router on the Internet, but
only the ones seen in the CAIDA traceroutes. Furthermore, we have geolocated every IP address found in the RIPE Atlas traceroutes. 
Due to CAIDAs measurement process and the monetary costs of geolocating large amounts of IP addresses, we were not able to geolocate every destination where CAIDA targeted a traceroute. 
Instead, we have geolocated a single IP address in every /26 we saw a traceroute to in CAIDAs measurements. 
This allows us to lookup IP addresses we do not have specific geolocation for using a longest matching prefix structure. 

\begin{figure}
    \centering
    \includegraphics[scale=.40]{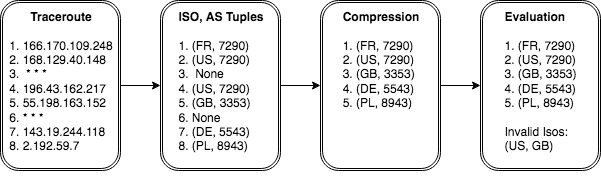}
    \caption{The process of transforming and formulating a traceroute based on its geographically (il)logical state.}
    \vspace{-5pt}
    \label{fig:traceroute-process}
\end{figure}

Processing a traceroute, visualized in
Figure~\ref{fig:traceroute-process}, involves converting an IP level
path into an aggregate path of (mapped country, AS tuples). When building this new path, we \emph{do not} add any hops from the
original traceroute measurement where we do not know the AS or country
that the IP address belongs to, nor do we include hops which do not
respond with an ICMP message to the traceroute. This implies that our
measurements and resulting analysis are a \emph{lower bound on the countries and ASes that the
path exposes traffic to}. This is critical to understanding that the raw
exposure we will soon present could potentially be \emph{worse} in practice.
Next, We compress repeated instances of the same tuple down to a single
instance. Finally, we label paths as either geographically normal or
``non-normal'' based on sets of expected countries to appear on the path constructed in Section~\ref{ssec:geoLogicConstruction}.

\subsection{Dataset Overview} \label{sec:dataset}

\begin{table}[h!]
\centering
\caption{An overview of our measurement results.}
\label{table:overview}
\resizebox{\columnwidth}{!}{%
\begin{tabular}{|l|c|c|}
\hline
Measurement Framework & \textit{CAIDA} & \textit{RIPE Atlas}\\ \hline
Measurement Goal & \textit{All-To-All Topology} & \textit{User Experience} \\ \hline
\# of Vantage Points & 174 & 76 \\ \hline
\# of ASes with Vantage Point(s) & 148 & 76  \\ \hline
\# of Countries with Vantage Point(s) & 52 & 30 \\ \hline
\# of Destination Countries Seen & 240 & 64 \\ \hline
\# of Paths Examined & 2,513,603,233 & 81,288  \\ \hline
Global Physical DoN & 0.565 & 0.632 \\ \hline
Global Legal DoN & 0.712 & 0.675 \\ \hline
\end{tabular}%
}
\end{table}

All together, we examined over 2.5 billion paths from both a physical and legal standpoint using our convex-hull definitions of normal. The examined CAIDA traceroutes were conducted between the dates January 1, 2018 and May 1, 2018. The examined RIPE Atlas traceroutes were conducted in February and March of 2018. We start our examination by looking at a general overview of the number and type of results we have obtained. Table \ref{table:overview} gives a brief overview of the quantity and type of measurements conducted. We see immediately that we have an incredibly large number of paths for our all-to-all topology measurement and a lesser number of paths for our user experience topology measurement. This is simply due to the nature of the measurement systems: CAIDA provides free traceroutes for us to use while the RIPE Atlas "on demand" system means we have to choose traceroutes wisely based on credits available and responsive probes.

\begin{figure}[]
\centering
\begin{minipage}[t]{0.75\columnwidth}
\centering
  \includegraphics[width=\columnwidth]{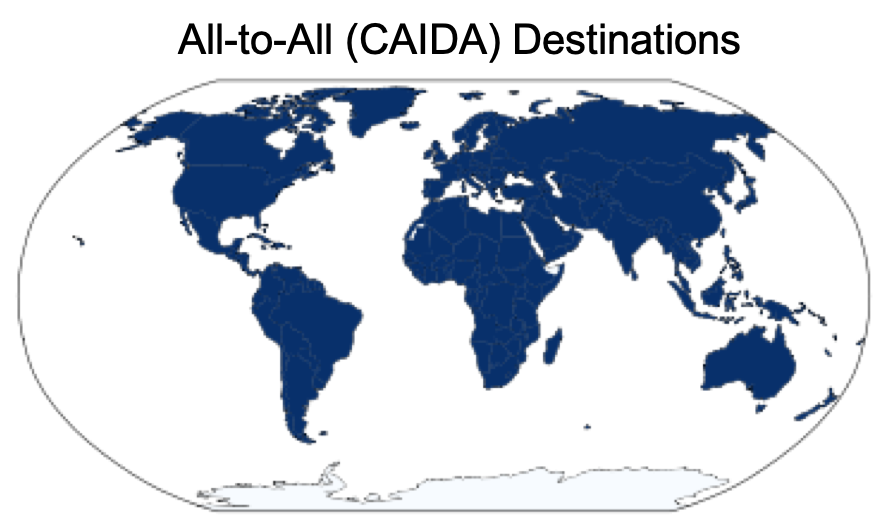}
  \caption{CAIDA's measurement structure gives us uniform access to destinations, which we see above.}
    \label{fig:caida-destinations}
\end{minipage}%
\hfill
\begin{minipage}[t]{0.75\columnwidth}
\centering
 \includegraphics[width=\columnwidth]{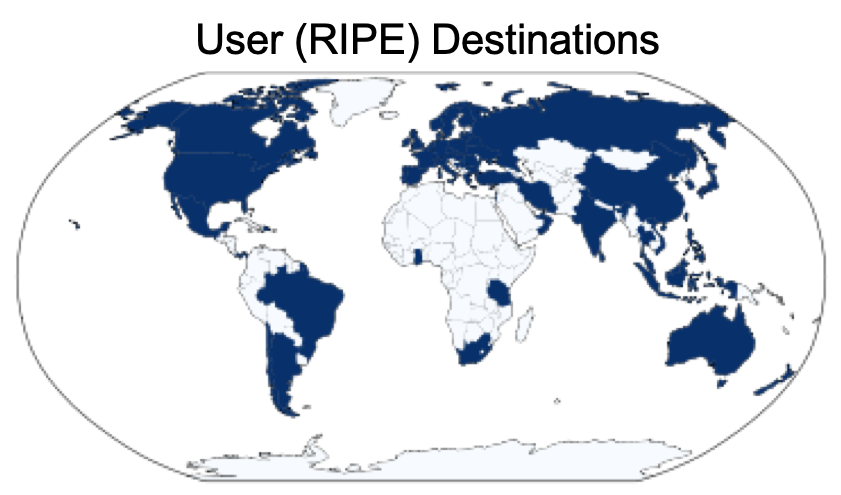}
  \caption{We see destinations on all major populated continents in the user study conducted via RIPE Atlas, but not all countries.}
   \label{fig:ripe-destinations}
\end{minipage}
\vspace{-5pt}
\end{figure}

CAIDA's measurement infrastructure conducts a traceroute to every announced /24 on the Internet, meaning we expect to see every country represented on this list, as there is \emph{some} amount of Internet infrastructure in every nation. On the other hand, our measurements from RIPE Atlas were specifically to Alexa Top 1000 websites, which are expected to be hosted closer to the majority of end users via CDNs, resulting in a concentration of these destinations in nations with more Internet users. Figure~\ref{fig:ripe-destinations} shows that we still see destinations on every major continent in the RIPE measurement data, but not every country.

\begin{figure}[]
\centering
\begin{minipage}[t]{0.95\columnwidth}
\centering
  \centering
 \includegraphics[width=0.70\columnwidth]{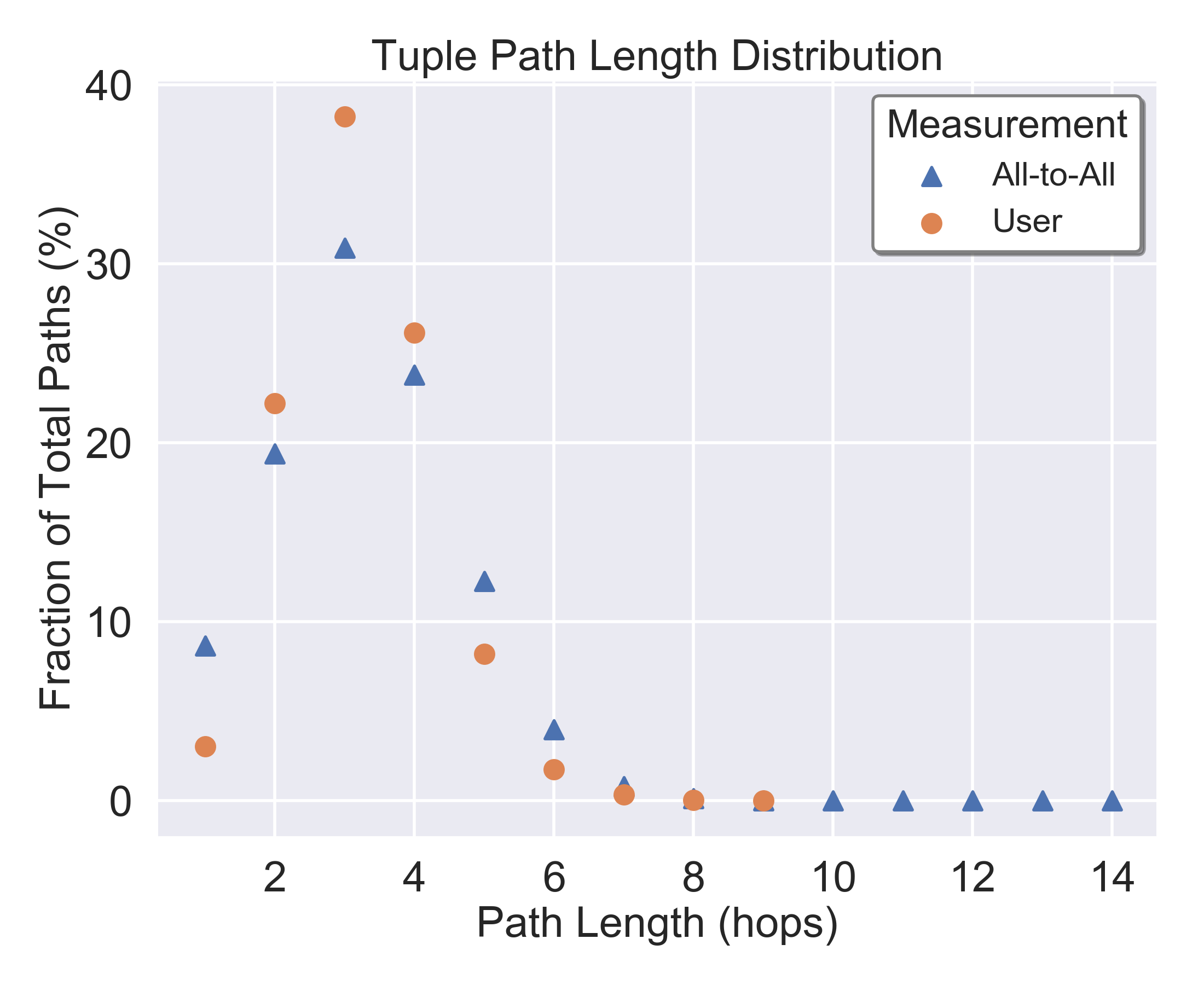}
  \caption{From a AS, Country tuple standpoint, the All-to-All paths are longer than the User study.}
\label{fig:tuple-path-length}
\end{minipage}%
\hfill
\begin{minipage}[t]{0.95\columnwidth}
  \centering
  \includegraphics[width=0.70\columnwidth]{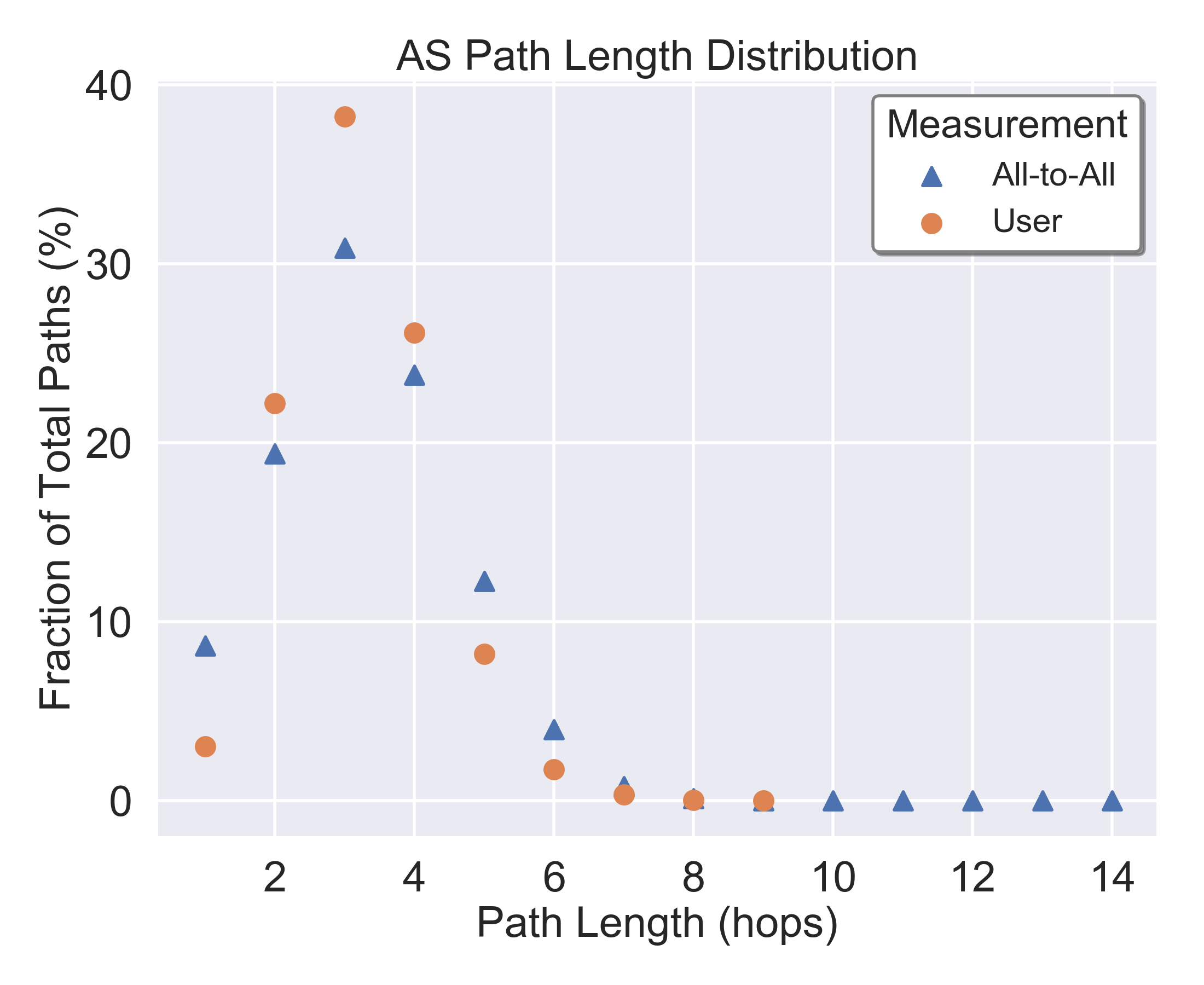}
  \caption{When considering the length of the path in terms of number of ASes, the All-to-All paths are still longer than the User study.}
    \label{fig:asn-path-length}
\end{minipage}
\vspace{-5pt}
\end{figure}

On average, we see that paths to CDNs are slightly shorter in terms of AS/country pairs than paths in our all-to-all dataset, as seen in Figure \ref{fig:tuple-path-length}. This trend is slightly less assertive in Figure \ref{fig:asn-path-length} where we see that at an AS level, the distribution of all-to-all measurements and user experience measurements are relatively close to each other. 

For the rest of the paper, we consider the datasets on the following levels: a country level, a regional level, and case studies of particularly interesting countries. Section~\ref{sec:physicalExposure} examines traffic exposure to different entities at a physical level: what countries the packets actually land in. Section~\ref{sec:legalExposure} then examines traffic exposure to different countries from a legal standpoint: where countries may have access to data that is not physically traversing them but through legal jurisdiction may have some access to the data.

\section{Physical Exposure} \label{sec:physicalExposure}

We want to quantify the amount of needless geographic exposure from
one entity to another. As a metric of normalcy, we have defined the
{\bf degree of normality (DoN)} between a particular source and
destination as:

\begin{equation}
DoN = \frac{total\ "normal\ paths"\ seen\ }{total\ paths\ seen}
\end{equation} \label{eq:don}

In this Section we consider only physical exposure of traffic, in Section~\ref{sec:legalExposure} we explore
what occurs when we additionally consider legal pressure nation
states can exert. The global physical DoN for the entire population of CAIDA or All-to-All measured paths is \textbf{0.565}. The global physical DoN for the entire population of RIPE Atlas or User Experienced paths is \textbf{0.632}, a roughly 11.9\% increase.

\begin{figure*}[!ht]
	\centering
	\subfloat[Tuple Path Length DoN]{\includegraphics[width=0.31\textwidth]{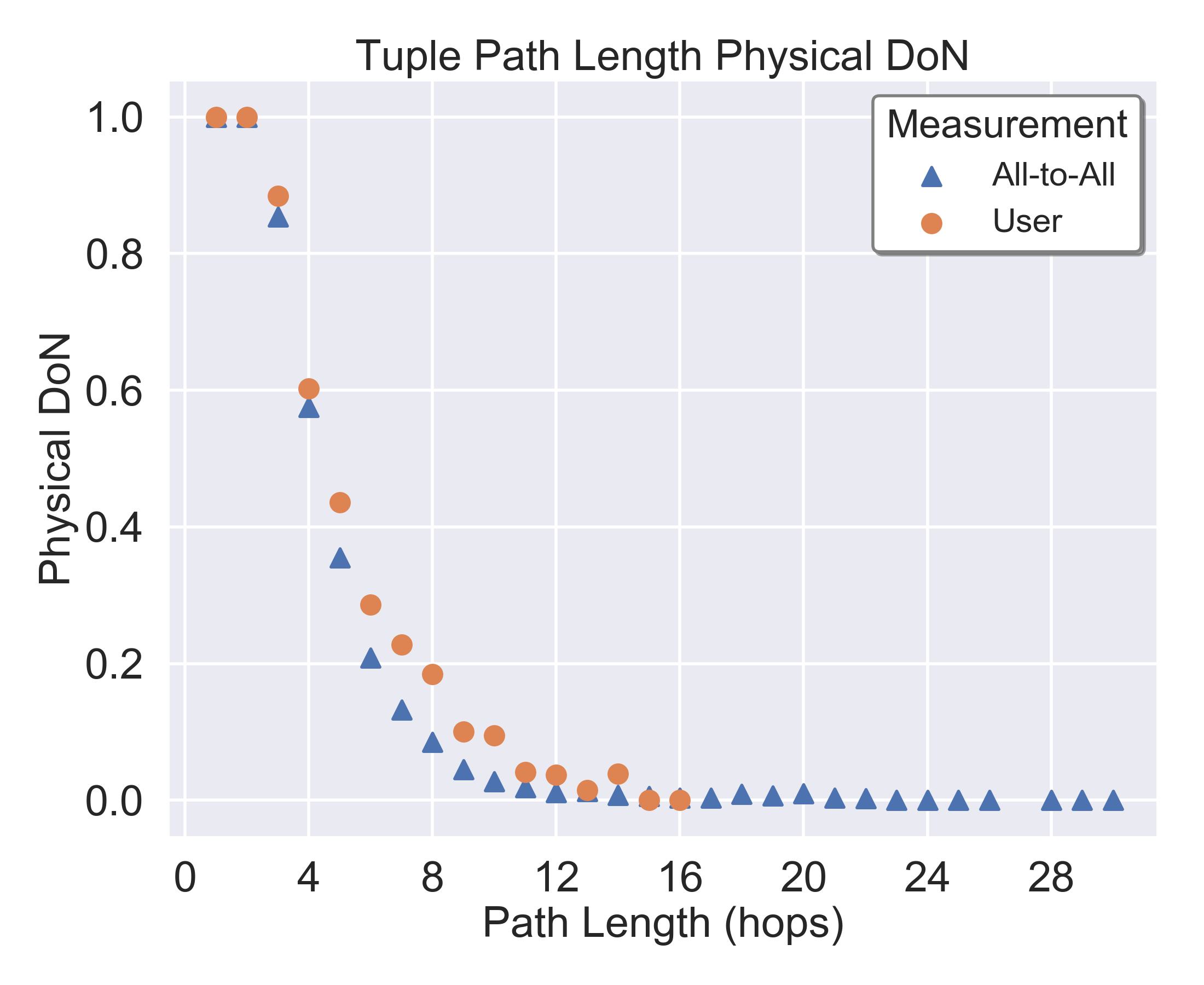}}
	\hspace{0.35cm}
	\subfloat[AS Path Length DoN]{\includegraphics[width=0.31\textwidth]{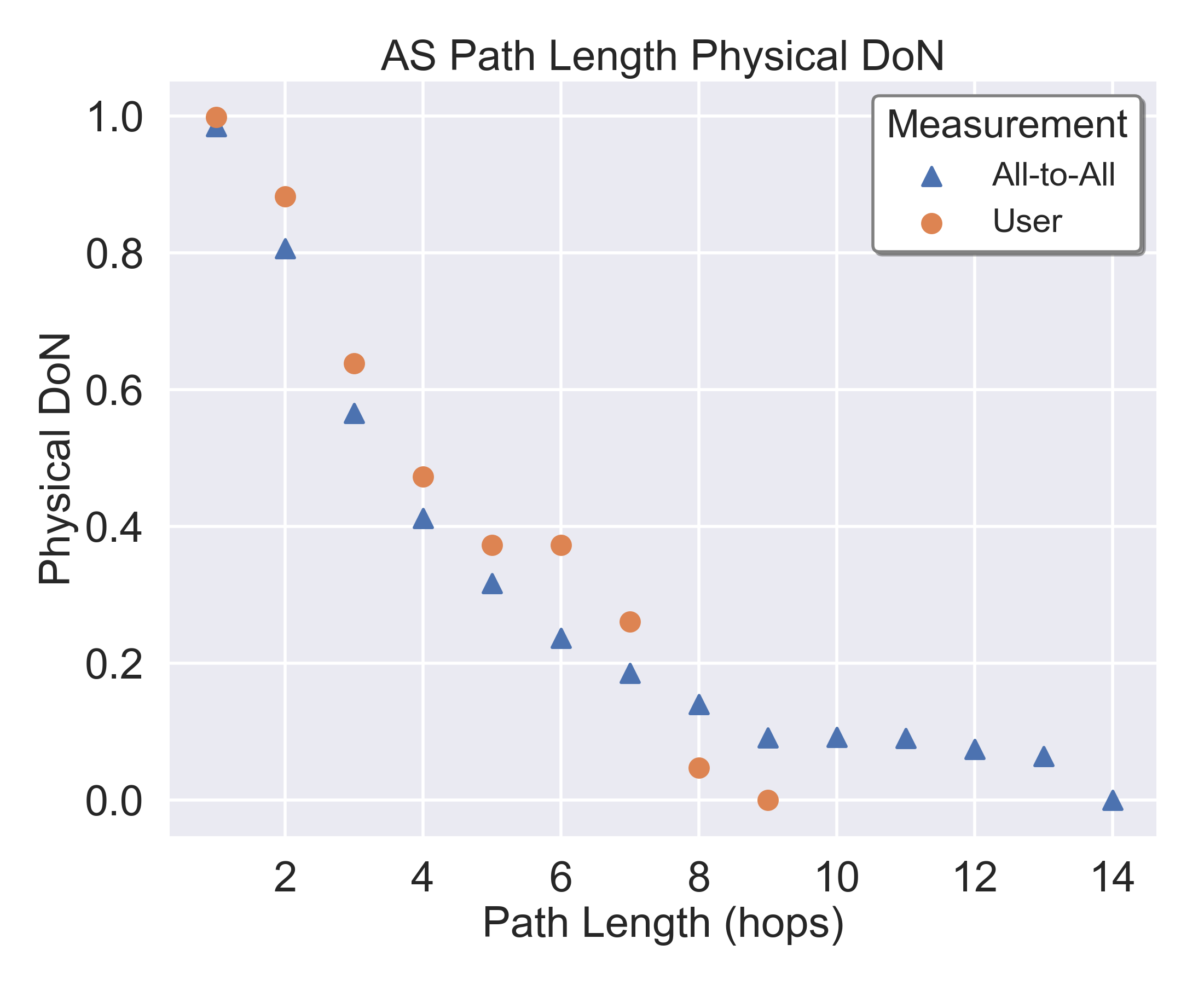}}
	\hspace{0.35cm}
	\subfloat[Severity of Irregular Paths]{\includegraphics[width=0.31\textwidth]{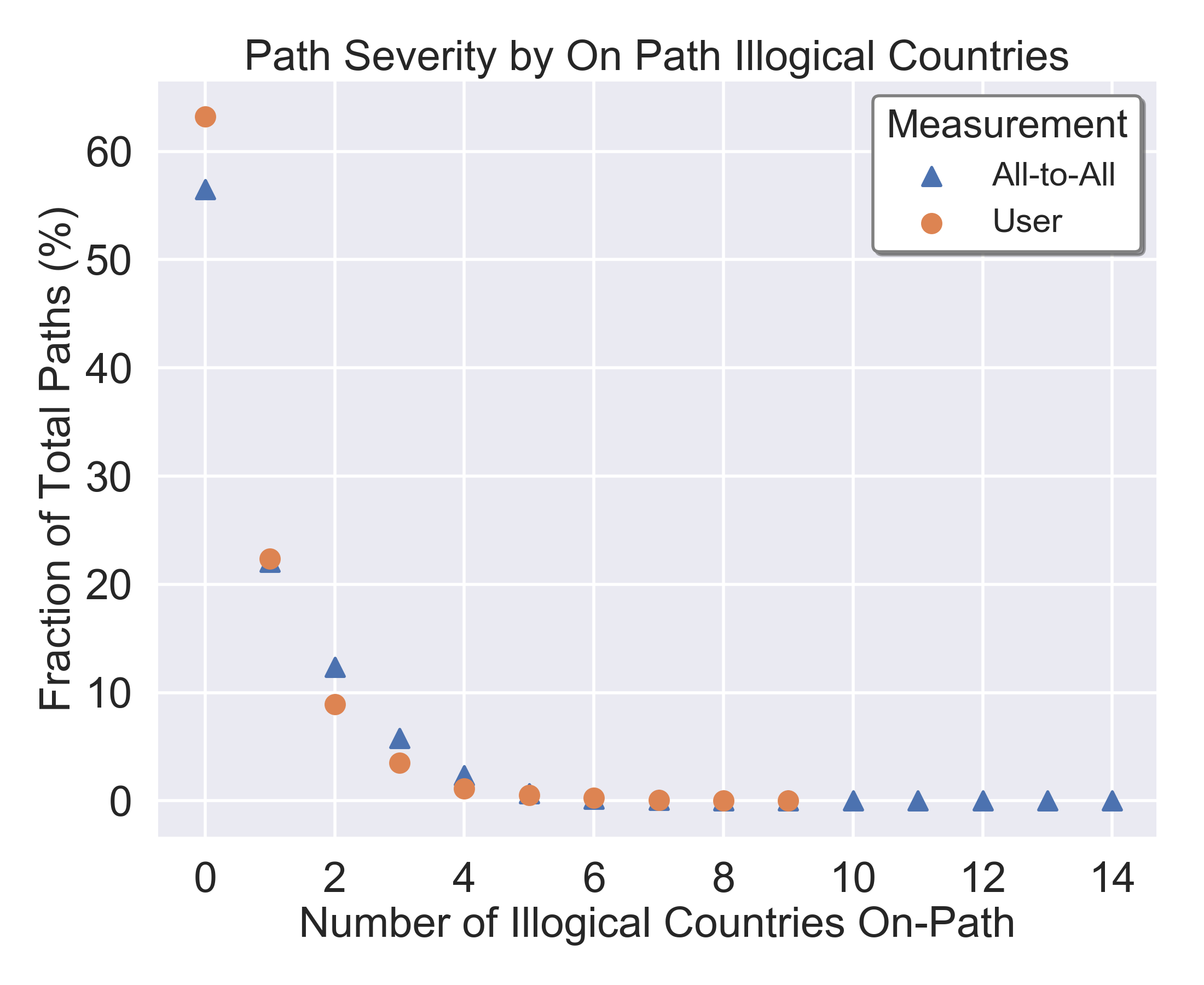}}
	\begin{center}
		\caption{Examining the DoN through different lenses.}{\label{fig:don-general}}
	\end{center}
	\vspace{-10pt}
\end{figure*}

We start our investigation of DoN by looking at the entirety of the measurements from a few different angles. First, we examine DoN
given the length of the compressed AS, country tuple path.
Figure~\ref{fig:don-general} shows that as the number of hops in the
path increases, the the DoN continually degrades. Only paths with less than 5
hops have a DoN above the global average for both all-to-all and user measurements, with DoN dropping below 0.2 when there are 8 or more AS, country hops in the paths. We also see that the longest path in the RIPE Atlas measurements is 16 hops, though we see paths of up to length 30 in the examined CAIDA traceroutes.

Beyond looking at just the raw number of AS, Country tuple hops in the
path, we investigated the DoN as the number of ASes involved grows. Figure~\ref{fig:don-general}b shows that similar results as the tuple length DoN. As the path grows longer in the number of ASes that carry the traffic, the less likely it is to be normal. Paths that contain more than 3 ASes in them have a DoN of below the global average for both datasets.

Lastly, Figure~\ref{fig:don-general}c gives us insight into the number
of geographically illogical countries appearing on each wrong path
that was observed. We term any country that is not geographically
normal between sender and receiver, but appears in a transit capacity
along the path a {\bf benefactor}. We see that the majority of
paths expose data to one or two benefactors. However, roughly a quarter of paths expose to traffic to 3 or more benefactors. Keep in mind that this is only one possible metric
of path error severity. One might also wish to consider
``intangibles'' related to the relationship between, source,
destination, and benefactor nations. For instance, one country may
want to avoid exposing internet traffic to specific countries for
political reasons. Another potential metric for path error severity is distance
from the built convex hull.

We split the rest of our examination of DoN into two levels: country and regional. Additionally, we split scenarios for each of the two entities based on the role of the entity in the path: the data source, the destination, or neither the source nor destination (a transit entity).

\subsection{Country DoN}   

\begin{figure}[]
\centering
\begin{minipage}[t]{0.95\columnwidth}
 \centering
    \includegraphics[width=0.80\columnwidth]{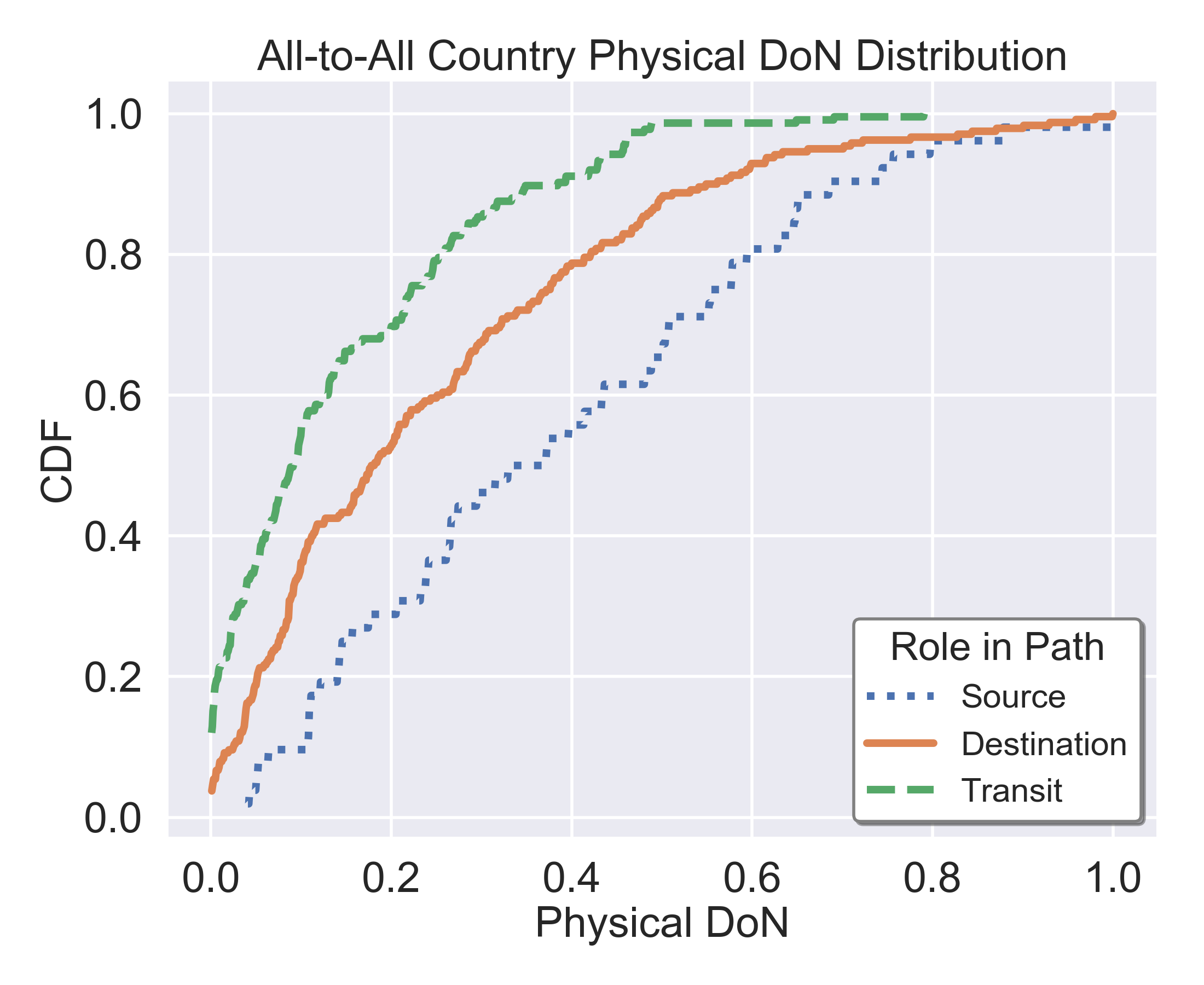}
    \caption{All-to-All Distribution of Physical DoNs}
    \label{fig:caida-country-don}
\end{minipage}%
\hfill
\begin{minipage}[t]{0.95\columnwidth}
    \centering
     \includegraphics[width=0.80\columnwidth]{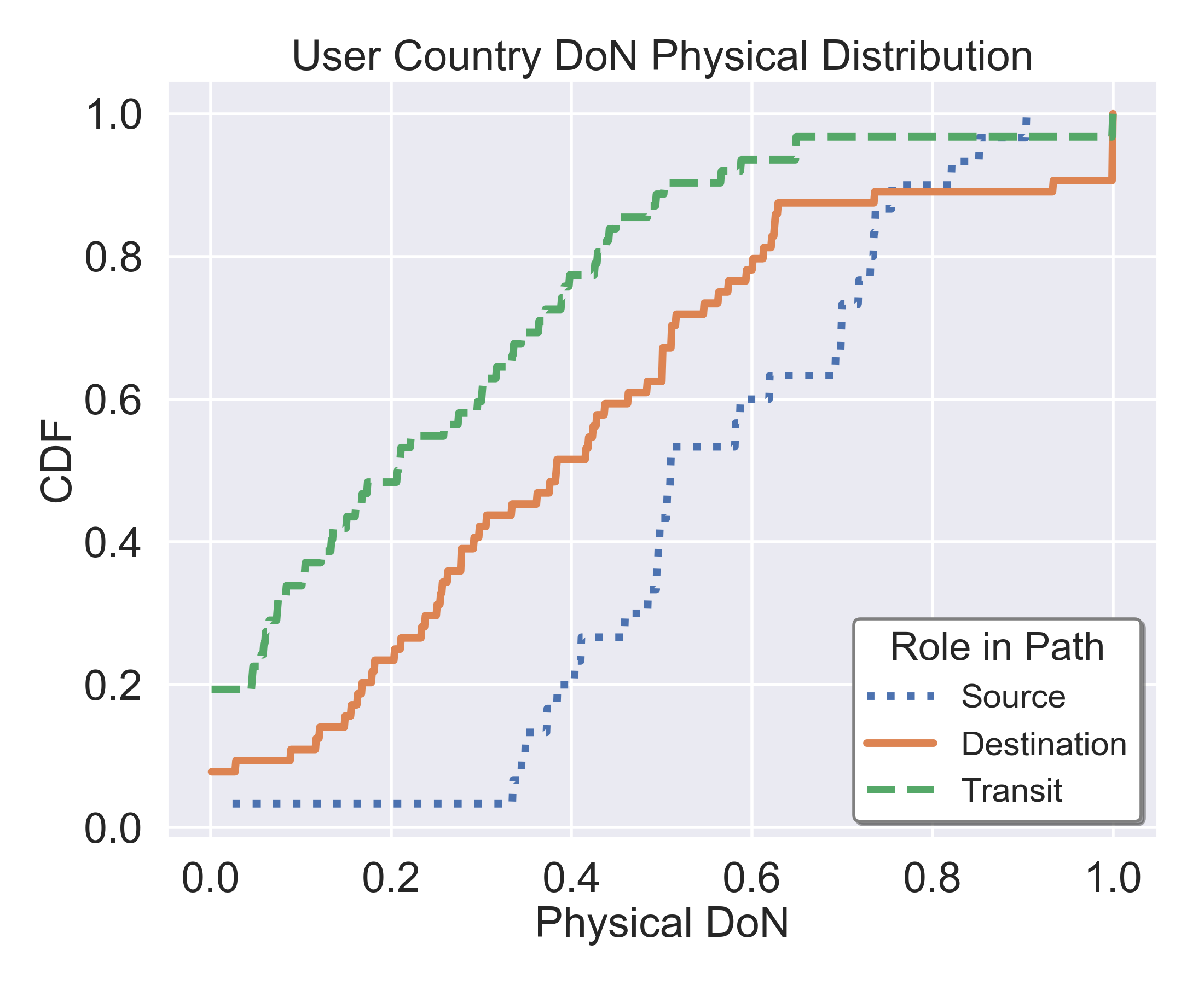}
    \caption{User Distribution of Physical DoNs}
    \label{fig:ripe-country-don}
\end{minipage}
\vspace{-5pt}
\end{figure}

DoN gives us the ability to examine how normal a population of paths is. Here we examine the DoN at a country level.
Figures~\ref{fig:caida-country-don} and~\ref{fig:ripe-country-don}
shows CDFs of the DoN of countries given their role in a
path. Immediately, we see that the DoN for each role in the path generally follows the same curve, but is shifted left. Particularly interesting is that in \emph{both} datasets transit DoN is by far the worst, while source DoN is better in every case. We see certain instances of a DoN of 0.00 or 1.00 in our RIPE measurements, these outliers are the result of exceptionally small numbers of observed paths involving the country in question.

While the global DoN for both measurements is above 0.500, we can see from the CDF that
this average is biased by a minority of countries, most notably the United
States, which appears both as the both the source and destination of a large
portion of its traffic. Only
roughly a quarter of countries have at or above the global DoN average. 

Expanding our examination into country DoN, we look into what countries serve as a transit provider for the most paths in our dataset, and how often such paths are geographically logical. Tables~\ref{table:caida-transit} and~\ref{table:ripe-transit} show the massive percentage of traffic that the United States transits in both of our datasets. The United States appears on more than half of paths as a transit provider, yet belongs on less than half of them. Also of particular interest is that Germany transits twice the percentage of traffic in the user measurements as it does in the all-to-all measurement, but also its transit DoN is also doubled when considering the user measurement. In fact, many of the European countries that transit the most traffic have significantly better transit DoNs in the user measurement than in the all-to-all measurement, reflecting the ability of CDNs to direct users to local copies of content, and as a side benefit cut down on needless exposure.

\begin{table*}[ht!]
\centering
\begin{minipage}[t]{.31\textwidth}
\centering
  \resizebox{1.0\columnwidth}{!}{%
    \begin{tabular}{|l|l|l|}
    \hline
    Country & Ratio of All Paths Transited & Transit DoN         \\ \hline
    US      & 0.537           & 0.480  \\ \hline
    GB      & 0.233          & 0.125  \\ \hline
    FR      & 0.112         & 0.141 \\ \hline
    DE      & 0.104          & 0.234 \\ \hline
    NL      & 0.091         & 0.265 \\ \hline
    ES      & 0.0791          & 0.077 \\ \hline
    CA      & 0.075          & 0.455 \\ \hline
    HK      & 0.053         & 0.053 \\ \hline
    ZA      & 0.039          & 0.087 \\ \hline
    SG      & 0.0388         & 0.130 \\ \hline
    \end{tabular}%
    }
    \caption{All-to-All Transit Providers} 
    \label{table:caida-transit}
\end{minipage}%
\hfill
\begin{minipage}[t]{.31\textwidth}
\centering
  \resizebox{1.0\columnwidth}{!}{%
\begin{tabular}{|l|l|l|}
\hline
Country & Ratio of All Paths Transited & Transit DoN          \\ \hline
US      & 0.365           & 0.449  \\ \hline
DE      & 0.234        & 0.493  \\ \hline
GB      & 0.207         & 0.301      \\ \hline
NL      & 0.164          & 0.438   \\ \hline
FR      & 0.131         & 0.441   \\ \hline
RU      & 0.117          & 0.565   \\ \hline
SE      & 0.069          & 0.274  \\ \hline
CA      & 0.065         & 0.335   \\ \hline
ES      & 0.062        & 0.046 \\ \hline
HK      & 0.043         & 0.167  \\ \hline
\end{tabular}%
}
\caption{User Transit Providers} 
\label{table:ripe-transit}
\end{minipage}%
\hfill
\begin{minipage}[t]{.31\textwidth}
\centering
  \resizebox{1.0\columnwidth}{!}{%
\begin{tabular}{|l|l|l|}
\hline
Country & Transit Only / Total Paths & Transit Only DoN \\ \hline
GB & 0.204 & 0.064 \\ \hline
US & 0.136 & 0.096 \\ \hline
FR & 0.092 & 0.092 \\ \hline
DE & 0.087 & 0.169 \\ \hline
NL & 0.069 & 0.160 \\ \hline
ES & 0.063 & 0.042 \\ \hline
HK & 0.050 & 0.034 \\ \hline
JP & 0.024 & 0.050 \\ \hline
KE & 0.023 & 0.101 \\ \hline
CA & 0.022 & 0.021 \\ \hline
\end{tabular}%
}
\caption{All-to-All Transit Only}
\label{table:caida-transit-only}
\end{minipage}%
\hfill
\begin{minipage}[t]{.31\textwidth}
\centering
\resizebox{0.99\columnwidth}{!}{%
\begin{tabular}{|l|l|l|}
\hline
Country & Transit Only / Total Paths & Transit Only DoN \\ \hline
GB & 0.179 & 0.244 \\ \hline
DE & 0.120 & 0.323 \\ \hline
US & 0.090 & 0.083 \\ \hline
NL & 0.079 & 0.279 \\ \hline
FR & 0.060 & 0.310 \\ \hline
ES & 0.059 & 0.029 \\ \hline
SE & 0.054 & 0.254 \\ \hline
CA & 0.042 & 0.065 \\ \hline
NO & 0.021 & 0.382 \\ \hline
HK & 0.020 & 0.002 \\ \hline
\end{tabular}%
}
\caption{User Transit Only}
\label{table:ripe-transit-only}
\end{minipage}%
\hfill
\begin{minipage}[t]{.33\textwidth}
\centering
 \resizebox{0.58\columnwidth}{!}{%
\begin{tabular}{|l|l|}
\hline
Country & \# Paths Benefited From \\ \hline
GB      & 457,562,290               \\ \hline
US      & 276,827,580               \\ \hline
FR      & 157,935,629               \\ \hline
DE      & 150,325,940               \\ \hline
NL      & 136,078,286               \\ \hline
ES      & 134,843,093               \\ \hline
HK      & 117,614,775               \\ \hline
CA      & 48,765,568                \\ \hline
JP      & 47,694,457                \\ \hline
KE      & 39,932,587                \\ \hline
\end{tabular}%
}
\caption{Top All-to-All Benefactors}
\label{table:caida-benefactors}
\end{minipage}%
\hfill
\begin{minipage}[t]{.31\textwidth}
\centering
\resizebox{0.62\columnwidth}{!}{%
\begin{tabular}{|l|l|}
\hline
Country & \# Paths Benefited From \\ \hline
GB      & 9,288                    \\ \hline
US      & 6,488                    \\ \hline
ES      & 4,242                    \\ \hline
DE      & 3,793                    \\ \hline
NL      & 3,580                    \\ \hline
CA      & 2,915                    \\ \hline
SE      & 2,615                    \\ \hline
FR      & 2,204                    \\ \hline
HK      & 1,591                    \\ \hline
JP      & 1,156                    \\ \hline
\end{tabular}%
}
\caption{Top User Benefactors}
\label{table:ripe-benefactors}
\end{minipage}
\vspace{-15pt}
\end{table*}

While looking at the magnitude of traffic each nation transits and its transit DoN is important, we realize that transit DoN can be inflated by paths that start and end in the same country. For instance, the United States contains multiple CAIDA monitors, and is the destination of many traceroutes, inflating its transit DoN. Perhaps more interesting is considering how geographically normal these countries are in paths that they are not the \emph{source} or \emph{destination} of. That is, they are \emph{only} a transit entity on the path. Tables~\ref{table:caida-transit-only} and~\ref{table:ripe-transit-only} show the vast difference when we focus on these paths. We now see that Great Britain gets to see an incredible amount of traffic that it is only a transit entity on, and only 6\% of that is normal. We also see that the United States gets to see much of the traffic it does because it is either the source or destination of that traffic. We once again see that many transit DoNs for countries are much higher in our user measurements.

In Figure~\ref{fig:don-general} we introduced the term 'benefactor' of a non-normal path. 
Tables~\ref{table:ripe-benefactors} and~\ref{table:caida-benefactors} show the countries that benefit the most from each set of paths examined. 
Immediately we see the extraordinary amount of paths that Great Britain and the United States see that they geographically should not. 
Great Britain is not only the top country in both measurements, but in the all-to-all measurement, Great Britain sees 3 times as many irregular paths as the 3rd most country. 
This is due in part to Great Britain's "piggybacking" off of bad paths bound for the United States, in addition to any illogical paths which traverse Great Britain by itself. 
Great Britain further benefits from paths originating from countries such as Spain and France "back tracking" to Great Britain before crossing the Atlantic Ocean.

\begin{figure*}[ht!]
\centering
\begin{minipage}[t]{0.45\textwidth}
\centering
\includegraphics[width=0.85\columnwidth]{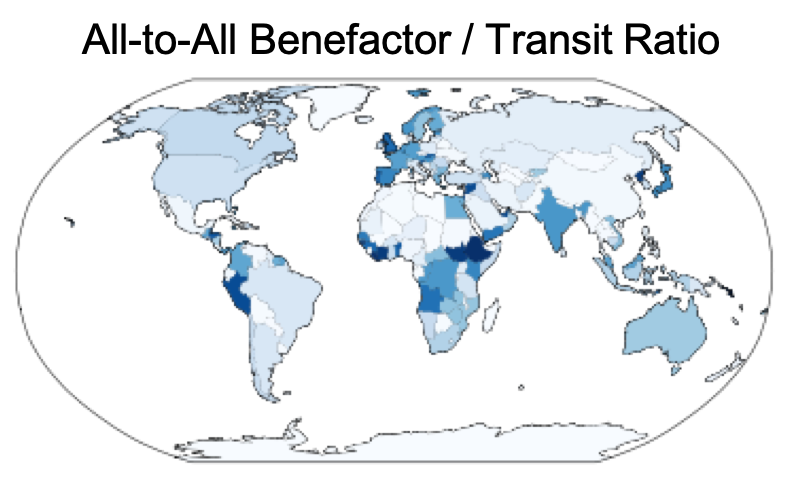}
    \caption{Geographic Distribution of the All-to-All (CAIDA) Benefactor/Transit Ratio}
    \label{fig:caida-ben-ratio}
\end{minipage}%
\hfill
\begin{minipage}[t]{0.45\textwidth}
    \centering
     \includegraphics[width=0.85\columnwidth]{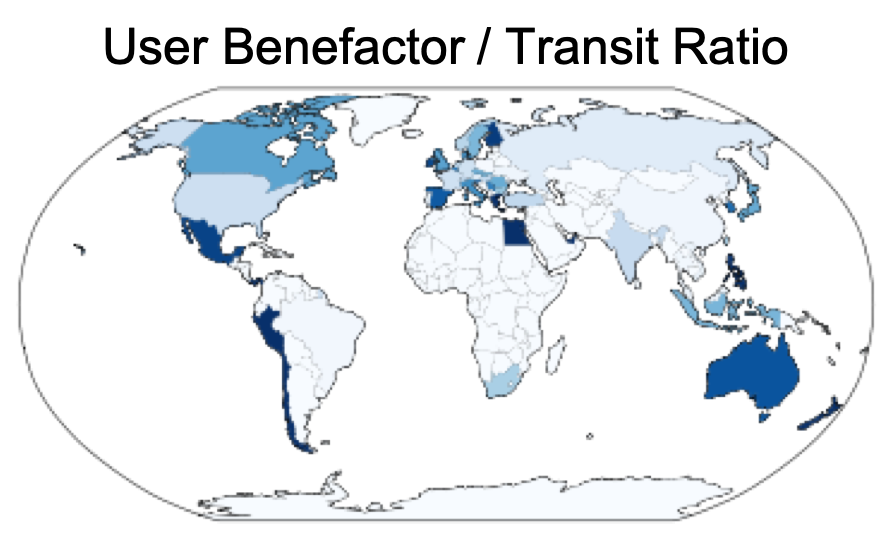}
    \caption{Geographic Distribution of the User (RIPE) Benefactor/Transit Ratio}
    \label{fig:ripe-ben-ratio}
\end{minipage}
\vspace{-5pt}
\end{figure*}

While these countries benefit from a large number of paths, we see that they do not necessarily do a bad job of transiting \emph{all} of their traffic. Magnitude is an issue with exposing traffic, but it is also valuable to examine the ratio of the number of paths a nation benefits from to the total amount of traffic the nation transits. Figures~\ref{fig:caida-ben-ratio} and \ref{fig:ripe-ben-ratio} visualize this ratio. We see that many of the top 10 countries from Tables~\ref{table:ripe-benefactors} and~\ref{table:ripe-benefactors} are not as prominent when considering this ratio. 
Instead, we see that for certain nations who appear as transit providers, in nearly all instances have no geographic business being on the path. In other words the vast majority of their transit traffic is a result of geographically illogical routes. One interesting point is that we see much of Africa benefits from large fractions of the traffic they transit in our all-to-all measurements but are virtually non-existent in the user measurements, a result of minimal CDN infrastructure appearing in the African continent.

\subsection{Regional DoN}

Next, we examine DoN at a regional level, where we see in
Tables~\ref{table:caida-region-don} and~\ref{table:ripe-region-don} that certain regions have a better
degree of normality when the path is to them rather than from them. For
instance, the Americas (North, South, and Central) have a higher than
average DoN when the path ends or starts there, but a much lower DoN
when they are found transiting data. Part of this could be explained
by the smaller number of countries in the Americas, particularly North
America. When having more adjacent countries, such as in Europe, there
are more choices of countries to route through, and could naturally
bring down the DoN for the region. Furthermore, we see that Africa's DoN is completely reversed when referencing both measurements. 
In the all-to-all measurement we see that they have a poor DoN when they are the
source or transit entity on a path, while in the user measurements they do a
much better job, but have poor destination DoN. Finally, we see that in Europe,
transit DoN rises dramatically in our user measurements, perhaps indicating that
they are a poor transit provider of traffic that is destined outside the region,
but have high levels of content localization in for destinations the Region. 

\begin{table}[ht!]
\centering
\begin{minipage}[t]{0.95\columnwidth}
\centering
\resizebox{0.85\columnwidth}{!}{%
\begin{tabular}{|l|l|l|l|}
\hline
Region   & Source DoN & Transit DoN & Destination DoN \\ \hline
Americas & 0.755      & 0.489       & 0.671           \\ \hline
Europe   & 0.455      & 0.228       & 0.521           \\ \hline
Asia     & 0.259      & 0.194       & 0.404           \\ \hline
Oceania  & 0.462      & 0.236       & 0.282           \\ \hline
Africa   & 0.154      & 0.116       & 0.347           \\ \hline
\end{tabular}%
}
\caption{All-to-All Regional DoN}
\label{table:caida-region-don}
\end{minipage}%
\hfill
\begin{minipage}[t]{0.95\columnwidth}
\centering
\resizebox{0.85\columnwidth}{!}{%
\begin{tabular}{|l|l|l|l|}
\hline
Region   & Source DoN & Transit DoN & Destination DoN \\ \hline
Americas & 0.739      & 0.454       & 0.719           \\ \hline
Europe   & 0.616      & 0.482       & 0.548           \\ \hline
Asia     & 0.386      & 0.178       & 0.247           \\ \hline
Oceania  & 0.565      & 0.064       & 0.570           \\ \hline
Africa   & 0.373      & 0.485       & 0.185           \\ \hline
\end{tabular}%
}
\caption{User Region DoN}
\label{table:ripe-region-don}
\end{minipage}
\vspace{-15pt}
\end{table}

Finally, we wish to examine the DoN of each region on a region to region basis. This is particularly important because it shows how normal the routing infrastructure is at a coarser level. Tables~\ref{table:caida-transit-to-region} and~\ref{table:ripe-transit-to-region} show that when staying inside a region, every region except Asia has an above average DoN. 
We also see in Table~\ref{table:caida-transit-to-region} that the DoN from one
region to another is highly symmetrical: the DoN traversing~\emph{from} region 1
to region 2 is typically close to the DoN when traversing from region 2
\emph{to} region 1. We see somewhat less symmetry in DoN in the user experience
dataset. Of particular note, Table~\ref{table:ripe-transit-to-region} reveals an incredibly high DoN from \emph{every} region to the Americas. 

\begin{table}[ht!]
\centering
\begin{minipage}[t]{0.95\columnwidth}
\centering
\resizebox{0.85\columnwidth}{!}{%
\begin{tabular}{|l|l|l|l|l|l|}
\hline
From \textbackslash To & Africa & Americas & Asia  & Europe & Oceania \\ \hline
Africa                 & 0.812  & 0.023    & 0.097 & 0.249  & 0.0003   \\ \hline
Americas               & 0.051  & 0.904    & 0.583 & 0.597  & 0.209   \\ \hline
Asia                   & 0.017  & 0.237    & 0.376 & 0.176  & 0.201   \\ \hline
Europe                 & 0.147  & 0.454    & 0.177 & 0.651  & 0.024   \\ \hline
Oceania                & 0.017  & 0.572    & 0.384 & 0.044  & 0.884   \\ \hline
\end{tabular}%
}
\caption{All-to-All Transit To/ From}
\label{table:caida-transit-to-region}
\end{minipage}%
\hfill
\begin{minipage}[t]{0.95\columnwidth}
\centering
\resizebox{0.85\columnwidth}{!}{%
\begin{tabular}{|l|l|l|l|l|l|}
\hline
From \textbackslash To & Africa & Americas & Asia  & Europe & Oceania \\ \hline
Africa                 & 1.000  & 0.535    & 0.050 & 0.186  & 0.000   \\ \hline
Americas               & 0.176  & 0.793    & 0.350 & 0.623  & 0.645   \\ \hline
Asia                   & 0.000  & 0.484    & 0.412 & 0.160  & 0.188   \\ \hline
Europe                 & 0.114  & 0.702    & 0.151 & 0.568  & 0.472   \\ \hline
Oceania                & 0.000  & 0.765    & 0.531 & 0.096  & 0.833   \\ \hline
\end{tabular}%
}
\caption{User Transit To/ From}
\label{table:ripe-transit-to-region}
\end{minipage}
\vspace{-15pt}
\end{table}

\section{Legal Exposure} \label{sec:legalExposure}

\begin{figure*}[ht!]
\centering
\begin{minipage}[t]{.32\textwidth}
    \includegraphics[width=1.0\columnwidth]{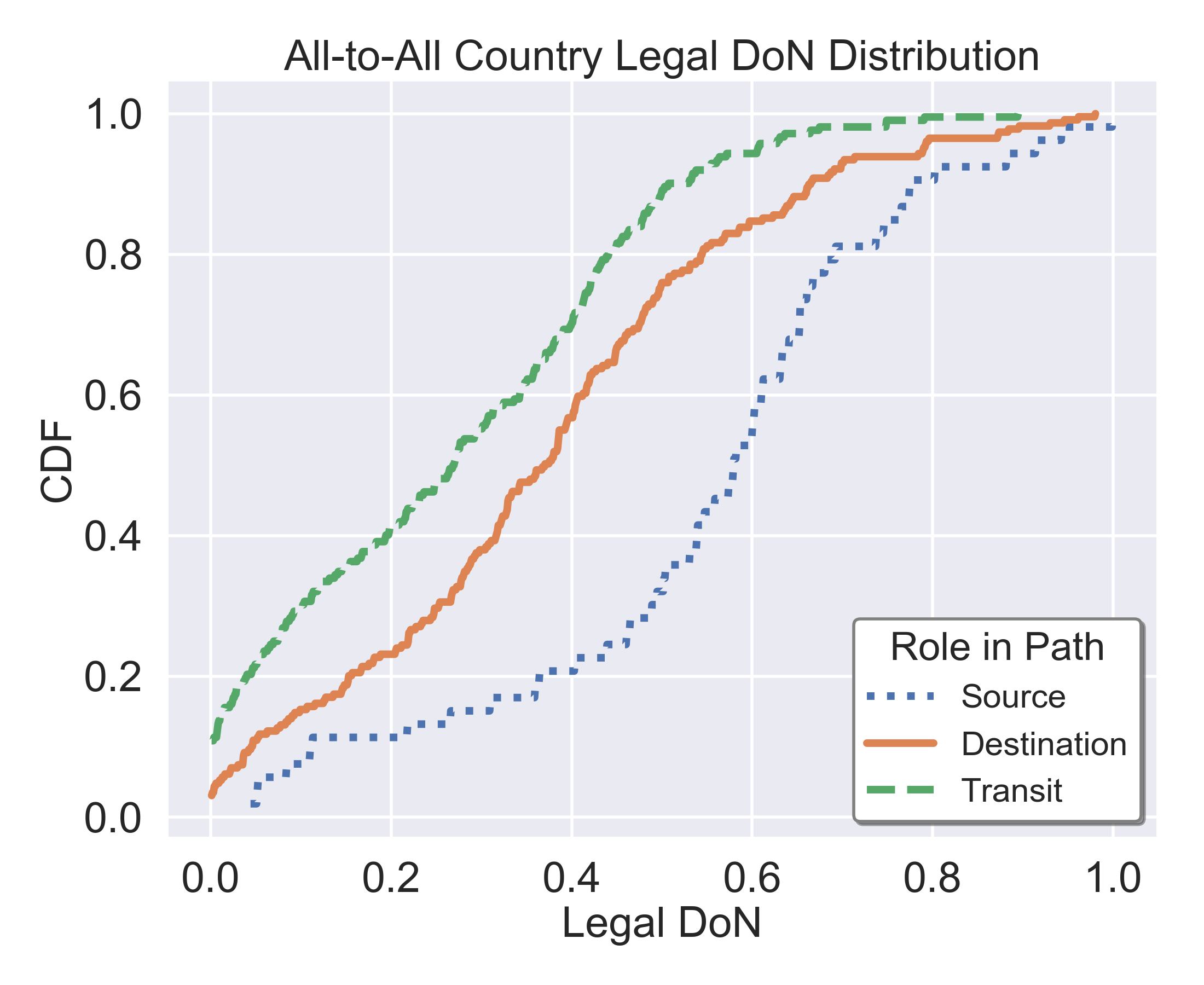}
    \caption{All-to-All Distribution of Legal Country DoNs}
    \label{fig:caida-country-don-legal}
\end{minipage}%
\hfill
\begin{minipage}[t]{.32\textwidth}
\includegraphics[width=1.0\columnwidth]{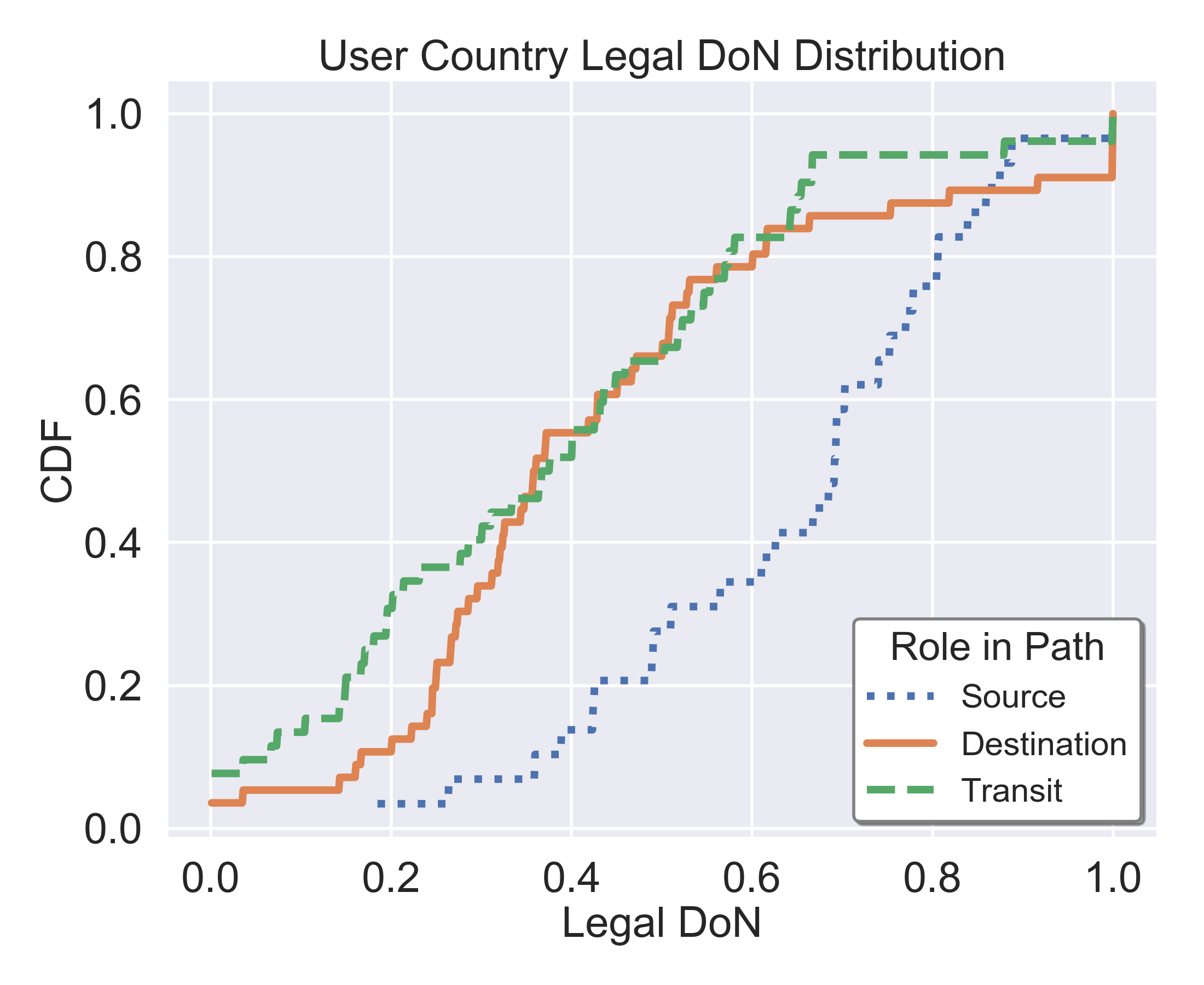}
    \caption{User Distribution of Legal Country DoNs}
    \label{fig:ripe-country-don-legal}
\end{minipage}%
\hfill
\begin{minipage}[t]{.32\textwidth}
\centering
     \includegraphics[width=1.0\columnwidth]{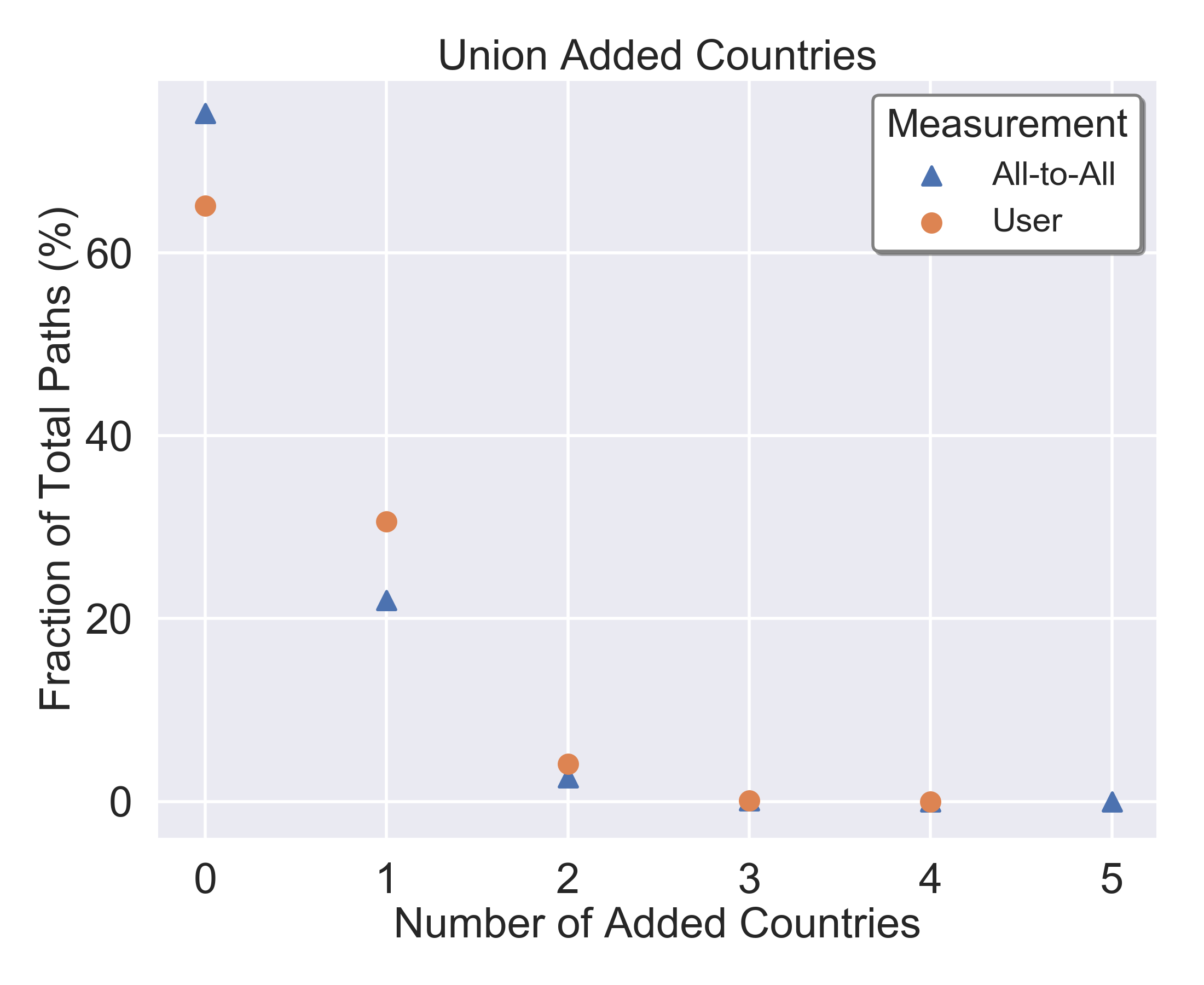}
    \caption{Added Countries when Unioning Legal and Physical Transit Countries}
    \label{fig:union-added-countries}
\end{minipage}
\end{figure*}

While looking at paths from a purely physical standpoint is important,
we also want to consider how 
geographically normal a path is when
additionally considering countries that can exert legal control over
an AS transiting traffic. This is particularly important because many ASes are legally registered in one country while having physical infrastructure in another. For example, 18 of our CAIDA monitors are physically located in a country different than the one their ASN is legally registered in. To measure the legal exposure of a path, we take every physical AS, country tuple path we have and map the AS back to the country it was
registered in. If we \emph{only consider legal exposure}, the global DoN for the
all-to-all measurement is 0.712 and the user measurement is 0.675. Figures~\ref{fig:caida-country-don-legal} and~\ref{fig:ripe-country-don-legal} show the legal DoN of the countries we have measurements for. We see mostly the same
trends here that we did in Figures~\ref{fig:caida-country-don} and~\ref{fig:ripe-country-don}, which is that the DoN of all three curves is mostly
the same, but shifted. One difference, however, is in the RIPE legal DoN. Here
we see higher transit DoNs than destination DoNs. Again we notice the trend that a small number of very well off nation states drive the overall average due to their involvement with more path. In fact, 80\% of nations have a DoN when involved in a path \emph{below} the average legal DoN.

\begin{table}[ht!]
\centering
\begin{minipage}[t]{0.70\columnwidth}
\centering
\resizebox{0.65\columnwidth}{!}{%
\begin{tabular}{|l|l|}
\hline
Country & \# Paths Benefited From \\ \hline
US      & 316,837,638               \\ \hline
BG      & 78,707,435                \\ \hline
DE      & 69,875,057                \\ \hline
HK      & 58,841,304                \\ \hline
GB      & 55,482,445                \\ \hline
ZA      & 46,602,480                \\ \hline
AT      & 28,613,961                \\ \hline
ES      & 27,396,710                \\ \hline
NL      & 26,234,852                \\ \hline
IL      & 21,173,866                \\ \hline
\end{tabular}%
}
\caption{All-to-All Legal Benefactors} 
\label{table:caida-legal-benefactors}
\end{minipage}%
\hfill
\begin{minipage}[t]{0.70\columnwidth}
\centering
\resizebox{0.65\columnwidth}{!}{%
\begin{tabular}{|l|l|}
\hline
Country & \# Paths Benefited From \\ \hline
BG      & 113,56                   \\ \hline
US      & 6,744                    \\ \hline
GB      & 2,888                    \\ \hline
DE      & 2,036                    \\ \hline
NL      & 1,291                    \\ \hline
UA      & 1,089                    \\ \hline
RU      & 1,059                    \\ \hline
DK      & 7,45                     \\ \hline
NO      & 6,95                     \\ \hline
ES      & 4,58                     \\ \hline
\end{tabular}%
}
\caption{User Legal Benefactors} 
\label{table:ripe-legal-benefactors}
\end{minipage}
\vspace{-15pt}
\end{table}

We also examine the benefactors of non-normal legal paths in Tables~\ref{table:caida-legal-benefactors} and~\ref{table:ripe-legal-benefactors}. Here we see Bulgaria, which was not in the top 10 benefactors for either measurement in terms of physical benefactors, is 1st and 2nd in our legal measurements of the all-to-all topology and user-experience, respectively. This is an exceptionally interesting case since few of the involved paths physically transit Bulgaria, but the Bulgarian government could place pressure on companies corporately headquartered there to gain access to the traffic.

Furthermore, we see that the United States is a benefactor in over 4 times as many paths as anyone else in our all-to-all measurements. This may be due to the fact that many large  ASes that are legally registered in the United States have physical infrastructure in multiple countries. 

\begin{table}[h]
\centering
\resizebox{0.40\columnwidth}{!}{%
\begin{tabular}{|l|l|l|}
\hline
 & All-to-All & User \\ \hline
Physical & 0.565 & 0.632 \\ \hline
Legal & 0.712 & 0.674 \\ \hline
Union & 0.519 & 0.500 \\ \hline
\end{tabular}%
}
\caption{Summary of the global DoN}
\label{table:don-summary}
\vspace{-10pt}
\end{table}

Ultimately, the path entities that have influence over the path that any data takes are \emph{both} the physical and legal entities that the traffic is exposed to. To examine the normality of paths when considering this, we have determined the normality of paths based on the union of the physical and legal countries transited. We term this DoN \emph{union DoN} and it accurately reflects all of the entities that \emph{could} act as path based adversaries on a given path. To compute this DoN we took the physical source and destination of a path and the union of both the legal and physical transit entities. Table~\ref{table:don-summary} shows the drop in DoN across both measurements when we compute this union. It is important to note that the union DoN can only be \emph{lower} than the physical DoN as we can only add countries to the set when computing the union of transit entities. We see a more than 20\% drop in normality for our user-experience data set when we consider the union of physical and legal exposure, as well as an 8\% drop in normality in our all-to-all measurement.

Figure~\ref{fig:union-added-countries} gives us insight into the amount of extra countries that a path gets exposed to when we compute the union of the physical and legal transit entities. Here we see that in most instances in which the union results in added exposure, only a single nation state is added.


\section{Related Work} \label{sec:related}

While our work is the first comprehensive study of which nation states inordinately benefit from choices made by the Internet's routing infrastructure, several other works have examined portions of the problem space. Karlin et. al.~\cite{karlin2009nation} was one of the first to explore this phenomena. In their work they utilized both traceroutes and inference based on information from BGP routing tables gathered for public mirrors to generate centrality of measures of nation states on the Internet. Our work builds on their techniques in order to explore if the centrality of certain nation states is simply the result of the number of hosts residing in those countries, if it results from providing a necessary physical connection between a send and receiver, or if, as we often saw, it is a physically unnecessary phenomena. Additionally, Karlin et. al. focused only on an "all-to-all" Internet measurement, rather than exploring the subset of paths used to distribute the majority of content.

Obar and Clement~\cite{obar2013internet} were the first to explore needless exposure of traffic to unnecessary nation states. Specifically they focused on paths taken from sources inside of Canada to destinations inside of Canada, but exited Canada temporarily to transit infrastructure in the United States. This "boomeranging" behavior they argued was a violation of the sovereignty of Canadian Internet traffic. Shah and Papadopoulos~\cite{shah2015characterizing} attempted to expand the measurement of this phenomena to all nations, basing their measurements again on observations of the global routing tables. Our work utilizes our novel convex hull based definition of geographically normal to more broadly explore this violation of sovereignty, expanding it to consider traffic transiting between two different countries in addition to traffic starting and ending inside the same nation.

The most similar to our work is Edmundson et. al.~\cite{edmundson2016characterizing}, which attempted to refine Shah and Papadopoulos's measurements of paths starting and ending inside the same nation transiting outside that nation, which they termed "tromboning". Like our work, Edmundson et. al. utilized traceroutes rather than inferences based on routing tables, and explored the concept of a "typical user" model. However, Edmundson et. al. focused their measurements on five countries: Brazil, Netherlands, Kenya, India, and the United States. Our measurements greatly expand on this; for the user model explored 30 and 64 countries as sources and destinations respectively, and our all-to-all model had 52 source and 240 destination nations. Additionally, their work utilized curl in order to build their model of which hosts are contacted during the loading of web content, a methodology that our prior work has found underestimates the number of resources contacted by a factor of roughly 20, and often incorrectly identifies the location of the server supplying the bulk of the content. Like similar work Edmundson et. al. utilize a focused definition of an illogical path, centered only on tromboning, while we apply our convex hull technique to explore geographically illogical paths more broadly. Lastly, given that their work pre-dated events such as the CLOUD act, Edmundson et. al. do not explore legal exposure in their work; given recent events we include this analysis.
 
Work exploring the degree to which other nation states can impact the Internet connectivity of their citizens or other nations as a result of the AS level Internet topology also exists. One of the most well known of these works is Roberts et. al.~\cite{roberts2011mapping}, where the authors explored how many ASes were physically located inside each country, and how many other nation states those ASes provided that country a direct connection to. More recently, Wahlisch et. al.~\cite{wahlisch2012exposing} explores this concept in detail for traffic sources or destinations specifically in the nation of Germany. Both of these works only consider what ASes have infrastructure inside or directly connected to the given nation states, and do not consider the actual paths traffic takes during transit and what nation states those paths traverse.

\newpage

\section{Conclusions} \label{sec:conc}

In this work, we have examined the extent to which the Internet's routing infrastructure
needlessly exposes network traffic to different nations. In summary, our contributions are as follows:
\begin{itemize}
\item We developed a  unique infrastructure for doing so by 
defining what a "normal" geographic path is between two countries through the use of convex hulls.
\item We quantified the amount of normal and irregular traffic between two entities through the lense of the path based degree of normality (DoN). We measure the exposure of both the entire geographic topology of the Internet and the subset of the topology that a user generally interacts with and examined over 2.5 billion paths.
\item We reveal that 44\% of paths concerning the entire geographic topology and 33\% of paths concerning the user experienced topology unnecessarily expose network traffic to at least one nation state, often more.
\item We examined the benefitting nations and regions of geographically illogical paths.
\item We explored the legal countries each measured path traverses to determine the countries with legal jurisdiction over the traversed ASes in all 2.5 billion paths. When considering both the physical and legal countries each path crosses, over 49\% of paths in both datasets expose traffic to at least one nation unnecessarily.
\item Overall, we found the global DoN for the physical topology to be 0.565 (entire topology) and 0.632 (user), and the legal to be 0.712 (entire) and 0.674 (user). The combined union of the two shows the entire Internet with a 0.519 DoN and user experienced paths to be 0.500. Notably, the Internet averages to be geographically logical in only slightly more than half of all countries and regions.
\end{itemize}

\noindent \textbf{Future Work:} We plan to expand our
measurements to examine countries which see temporary, but marked,
changes in their DoN. We will then attempt to establish the root cause of such
changes. Ideally, we could measure these changes in real-time via measurement infrastructures built into CAIDA systems such as Ark or RIPE Atlas. We are also interested in examining if adversarial actions could
result in a temporarily reduced DoN for nations, or if particular
nations could inordinately benefit from adversarial reductions in DoN.
Lastly, we wish to examine if nations can adjust their routing
policies in an effort to increase their DoN, effectively reducing
their exposure to nation state level path adversaries. 

\newpage


\end{document}